\documentclass[preprintnumbers, floatfix, letterpaper, aps, prd, epsfig, nofootinbib, twocolumn]{revtex4-1}
\usepackage{bm, graphicx, dcolumn, epstopdf, epsf, latexsym, mathbbol, amssymb, amsmath, color, slashed, mathrsfs, mathcomp, simplewick}
\usepackage[center]{subfigure}
\usepackage{multirow}
\usepackage{makecell}
\usepackage[colorlinks,linkcolor=blue,citecolor=blue,urlcolor=blue]{hyperref}

\begin{document}
\allowdisplaybreaks
 \newcommand{\bq}{\begin{equation}} 
 \newcommand{\eq}{\end{equation}}
 \newcommand{\bqn}{\begin{eqnarray}}
 \newcommand{\eqn}{\end{eqnarray}}
 \newcommand{\nb}{\nonumber}
 \newcommand{\lb}{\label}
 \newcommand{\f}{\frac}
 \newcommand{\p}{\partial}
\newcommand{\PRL}{Phys. Rev. Lett.}
\newcommand{\PLB}{Phys. Lett. B}
\newcommand{\PRD}{Phys. Rev. D}
\newcommand{\CQG}{Class. Quantum Grav.}
\newcommand{\JCAP}{J. Cosmol. Astropart. Phys.}
\newcommand{\JHEP}{J. High. Energy. Phys.}

\title{Gravitational radiations from periodic orbits around a black hole in the effective field theory extension of general relativity}

\author{Shuo Lu${}^{a, b}$}
\email{lushuo@zjut.edu.cn}

\author{Hao-Jie Lin${}^{c, d}$}
\email{haojielin@stumail.neu.edu.cn}

\author{Tao Zhu${}^{a, b}$}
\email{zhut05@zjut.edu.cn; Corresponding author}

\author{Yu-Xiao Liu${}^{e, f}$}
\email{liuyx@lzu.edu.cn}

\author{Xin Zhang${}^{c, d}$}
\email{zhangxin@mail.neu.edu.cn}

\affiliation{
${}^{a}$Institute for Theoretical Physics \& Cosmology, Zhejiang University of Technology, Hangzhou, 310023, China\\
${}^{b}$ United Center for Gravitational Wave Physics (UCGWP),  Zhejiang University of Technology, Hangzhou, 310023, China\\
${}^{c}$Key Laboratory of Cosmology and Astrophysics (Liaoning) \& College of Sciences,
Northeastern University, Shenyang 110819, China\\
${}^{c}$Key Laboratory of Data Analytics and Optimization for Smart Industry (Ministry of Education), Northeastern University, Shenyang 110819, China \\
${}^{e}$Lanzhou Center for Theoretical Physics, Key Laboratory of Theoretical Physics of Gansu Province, Lanzhou University, Lanzhou 730000, China\\
${}^{f}$Institute of Theoretical Physics \& Research Center of Gravitation, Lanzhou University, Lanzhou 730000, China}

\date{\today}

\begin{abstract}

The study of periodic orbits in extreme-mass-ratio inspirals is essential for understanding the dynamics of small bodies orbiting supermassive black holes. In this paper, we study the periodic orbits and their corresponding gravitational wave emissions within the framework of an effective field theory-based extension of general relativity (EFTGR), which incorporates higher-order curvature terms into the Einstein-Hilbert action. We start with a brief analysis of the modified black hole spacetime in EFTGR and examine how its parameters influence the dynamics of a massive neutral particle using the Lagrangian formalism. Focusing on the impact of the higher-order curvature terms in EFTGR, we examine the properties of periodic orbits, which are characterized by three topological integers $(z, w, v)$ that uniquely classify their trajectories. By analyzing these orbits within EFTGR, we aim to provide new insights into how strong-field deviations from general relativity may manifest in observable phenomena. We then calculate the gravitational waveforms generated by these periodic orbits, identifying potential observational signatures. Our analysis reveals a direct connection between the zoom-whirl orbital behavior of the small compact object and the gravitational waveforms it emits: higher zoom numbers lead to increasingly intricate waveform substructures. The results contribute to a clearer understanding of the dynamical features of EFTGR and open new avenues for probing black hole properties via gravitational wave detection.

\end{abstract}

\maketitle

\section{Introduction}
\renewcommand{\theequation}{1.\arabic{equation}} \setcounter{equation}{0}

With the pioneering detections by LIGO and Virgo \cite{LIGOScientific:2016aoc, LIGOScientific:2016vbw, LIGOScientific:2016vlm, LIGOScientific:2016emj}, gravitational wave (GW) astronomy opened a new observational window into the universe's most extreme phenomena, such as binary black holes and binary neutron star mergers. Nevertheless, general relativity (GR) continues to confront fundamental issues, including the singularity problem, which remains a topic of deep investigation \cite{Hawking:1970zqf, Senovilla:1998oua}. GW radiation from extreme mass-ratio inspirals (EMRIs) is particularly valuable in this regard, as it encodes detailed information about the strong-field region and serves as a promising tool for probing black hole properties.

For EMRIs, the two bodies have an extreme mass ratio, with the lighter companion executing rapid orbits around the central supermassive black hole. These systems are prime targets for future space-based gravitational wave missions, including Taiji \cite{Hu:2017mde}, Tianqin \cite{TianQin:2015yph, Gong:2021gvw, Luo:2025ewp}, and LISA \cite{Danzmann:1997hm, Schutz:1999xj, Gair:2004iv, LISA:2017pwj, Maselli:2021men}. Their long-lived signals encode detailed information about orbital dynamics and the background spacetime, enabling precision tests of strong-field gravity and cosmic evolution \cite{Bian:2025ifp, Ni:2024acg}. Given that the smaller body loses energy only gradually, the inspiral can last for years and be effectively described by a sequence of periodic orbits. The GWs emitted in this regime are sensitive to the underlying theory of gravity, making EMRIs powerful probes for testing deviations from GR, see refs. \cite{Kumar:2025jsi, Kumar:2025njz, Kumar:2024our, Kumar:2024utz, Zi:2024jla, AbhishekChowdhuri:2023gvu, Zhang:2023kzs} for examples. Observationally, they allow accurate measurements of supermassive black hole parameters such as mass, spin, and quadrupole moment \cite{Amaro-Seoane:2007osp}, while cosmologically they act as standard sirens that constrain key parameters \cite{Schutz:1986gp, Jin:2025dvf}. In addition, environmental effects, such as dynamical friction from surrounding dark matter, can leave subtle imprints on the EMRIs waveform, offering a means to investigate the distribution of matter near galactic centers  \cite{Zhang:2022roh, Zhang:2021bdr}. Since the inspiral dynamics can be effectively represented as a sequence of periodic trajectories, a systematic framework for classifying such orbits becomes essential.

To this end, a taxonomy has been introduced based on three integers: scaling ($z$), rotation ($w$), and vertex ($v$) \cite{Levin:2008mq}. This scheme provides a unified framework for classifying closed orbits and has been widely applied to diverse spacetimes, including Schwarzschild and Kerr black holes \cite{Levin:2008ci, Levin:2009sk, Bambhaniya:2020zno, Rana:2019bsn}, charged black hole \cite{Misra:2010pu}, naked singularities \cite{Babar:2017gsg}, Kerr-Sen black holes \cite{Liu:2018vea}, and hairy black holes in Horndeski gravity \cite{Lin:2023rmo}. For studies of periodic orbits in other black holes, see refs.~\cite{Yao:2023ziq, Lin:2022llz, Chan:2025ocy, Wang:2022tfo, Lin:2023eyd, Habibina:2022ztd, Zhang:2022psr, Lin:2022wda, Gao:2021arw, Lin:2021noq, Deng:2020yfm, Zhou:2020zys, Gao:2020wjz, Deng:2020hxw, Azreg-Ainou:2020bfl, Wei:2019zdf, Pugliese:2013xfa,Zhang:2022zox,Chan:2025ocy, Al-Badawi:2025yum,Healy:2009zm,Wei:2025qlh,Sharipov:2025yfw}.
The associated gravitational radiation from such periodic trajectories has also been investigated in many contexts \cite{Tu:2023xab, Yang:2024lmj, Shabbir:2025kqh, Junior:2024tmi, Zhao:2024exh, Jiang:2024cpe, Yang:2024cnd, Meng:2024cnq, Li:2024tld, QiQi:2024dwc, Lu:2025cxx, Haroon:2025rzx, Chen:2025aqh, Alloqulov:2025ucf, Choudhury:2025qsh, Wang:2025hla, Alloqulov:2025bxh, Li:2025sfe, Wang:2025wob, Gong:2025mne, Zare:2025aek, Zahra:2025tdo, Li:2025eln, Deng:2025wzz, Ahmed:2025azu, Alloqulov:2025dqi, Zhang:2025wni}. 

In this paper, we investigate the periodic orbital motions of a test particle and their GW emissions around a black hole within the framework of an effective field theory-based extension of general relativity (EFTGR). This theory modifies general relativity by adding higher-order curvature terms to the Einstein-Hilbert action \cite{Endlich:2017tqa}. EFTGR provides a systematic methodology to capture and parameterize all possible modifications to an existing theory that may emerge when new physics is introduced. Its construction is guided by a set of principles \cite{Endlich:2017tqa}: it must be (1) testable via gravitational-wave 
observations; (2) consistent with other experimental tests of general relativity, including short-distance probes; (3) compatible with fundamental principles such as locality, causality, and unitarity; and (4) devoid of new light degrees of freedom. Under these assumptions, a unique effective field theory framework has been formulated, exhibiting several attractive properties\cite{Endlich:2017tqa}. By studying the signatures of a single Lagrangian, a vast class of theories is covered, since any framework consistent with the stated assumptions will reproduce the same phenomenology as EFTGR under suitable parameter choices. Within this approach, all relevant information is compactly encoded in a small number of coefficients, which can be constrained through observations. Unlike purely phenomenological parametrizations, the formulation of EFTGR intrinsically restricts the investigation to physically consistent theories and, within this domain, provides an optimal general parametrization of the observational signals. In recent years, numerous studies related to black holes and GWs have been carried out in the framework of EFTGR, see refs. \cite{deRham:2020ejn, Cayuso:2020lca, Cano:2021myl, deRham:2021bll, Cano:2022wwo, Cao:2025qws, Bernard:2025dyh, Maenaut:2024oci, Cardoso:2018ptl, Cano:2024wzo, Liu:2024atc, Silva:2022srr, BarrosoVarela:2023ull} and references therein.

In the context of black hole physics, EFTGR corrections typically decay rapidly with radial distance, rendering the weak-field regime indistinguishable from GR. However, near the horizon and in the strong-field region, these corrections can become significant, modifying observables such as tidal Love numbers, quasinormal modes, and orbital dynamics \cite{Cardoso:2018ptl}. Periodic orbits of test particles around black holes are particularly sensitive probes of such corrections, as they encode detailed information about the background geometry and directly influence the GW signal. In the framework of EFTGR, several typical high-order curvature terms are added to the gravitational action. Among them,  a few certain terms can uniquely govern the corrections to the metric of the spherically symmetric static black hole spacetimes. This motivates our study, which aims to clarify how these corrections modify orbital motion and the resulting GW signatures.

The article is constructed as follows. In Sec.~\ref{section2}, we give a brief introduction to EFTGR and the spherical symmetric black hole solution. And then in Sec.~\ref {section3}, we discuss the geodesic of a massive test particle around the black hole in EFTGR. The effective potential and bounded orbits of the test particle are also discussed. In Sec.~\ref{section4}, a detailed classification of periodic orbits is provided. In Sec.~\ref{section5}, the GW radiation from certain periodic orbits around the black holes in EFTGR is presented. The conclusion and discussion are summarized in Sec.~\ref{section6}.

In this paper, we adopt the unit system by setting $c=G_N=1$, where $c$ denotes the speed of light and $G_N$ the Newtonian gravitational constant. We follow the standard convention: Greek indices $(\mu,\nu,...)$ run from $0$ to $3$, Latin indices $(i,j,k, ...)$ from $1$ to $3$, and the metric signature is taken as $(-,+,+,+)$ .

\section{Black hole in the effective field theory extension of general relativity}\label{section2}
\renewcommand{\theequation}{2.\arabic{equation}} \setcounter{equation}{0}

The action of EFTGR is expressed as
\bqn\label{action}
S_{\rm eff}=2M_{\rm Pl}^{2}\int d^{4}x\sqrt{-g}\left(R-\frac{C^{2}}{\Lambda^{6}}-\frac{\tilde{C}^{2}}{\tilde{\Lambda}^{6}}-\frac{C\tilde{C}}{\Lambda_{-}^{6}}\right) + \cdots,\nb\\
\eqn
where $M_{\rm Pl} = (8\pi G)^{-1/2}$ is the Planck mass with $G$ being the gravitational constant, $R$ is the Ricci scalar of the four-dimensional spacetime with metric determinant $g$. And
\bqn
C \equiv R_{\alpha \beta \gamma \delta}R^{\alpha \beta \gamma \delta},\quad \tilde{C}\equiv R_{\alpha \beta \gamma \delta}\epsilon^{\alpha \beta}_{\quad \mu \nu}R^{\mu \nu \gamma \delta},
\eqn
where $\tilde{R}^{\alpha\beta\gamma\delta}\equiv\epsilon^{\alpha\beta}_{\quad\mu\nu}R^{\mu\nu\gamma\delta}$ is the dual tensor with $R^{\mu \nu \gamma \delta}$ being the Riemann tensor, $\epsilon_{\mu\nu\rho\sigma}$ is the total antisymmtric Levi-Civita tensor defined such that $\epsilon^{0123}=1/\sqrt{-g}$, and $\Lambda$, $\tilde{\Lambda}$, $\Lambda_{-}$ are energy cutoff scales for each operator.

By varying the action (\ref{action}) with respect to the metric $g_{\mu\nu}$, one obtains the field equation of EFTGR, which reads as
\bqn
&&R^{\mu\alpha}-\frac{1}{2}g^{\mu\alpha}R\nb\\
&& =\frac{1}{\Lambda^{6}}\left(8R^{\mu\nu\alpha\beta} \nabla_{\nu} \nabla_{\beta}C + \frac{g^{\mu \alpha}}{2}C^{2} \right)\nb\\
&& \;\;\;+\frac{1}{\tilde{\Lambda}^{6}} \left(8\tilde{R}^{\mu\rho\alpha\nu}\nabla_{\rho}\nabla_{\nu}\tilde{C}+\frac{1}{2}g^{\mu\alpha}\tilde{C}^{2}\right)\nb\\
&&\;\;\; + \frac{1}{\Lambda^{6}_{-}} \left( 4\tilde{R}^{\mu\rho\alpha\nu} \nabla_{\rho}\nabla_{\nu}C+4R^{\mu\rho\alpha\nu}\nabla_{\rho}\nabla_{\nu}\tilde{C} +\frac{g^{\mu\alpha}}{2}\tilde{C}C \right).\nb\\
\eqn
The spherically symmetric black hole solution can be derived by solving the above field equation and has been studied in Ref.~\cite{Cardoso:2018ptl}. Consider a spherically symmetric, static vacuum solution,
\bqn
ds^2=-f^{\epsilon_i}_t(r)dt^2+\frac{1}{f^{\epsilon_i}_r(r)}dr^2+r^2d\Omega^2,
\eqn
where $d\Omega^2=d\theta^2+\sin^2{\theta}d\phi^2$ and the two metric functions $f^{\epsilon_i}_{t}(r)$ and $f^{\epsilon_i}_{r}(r)$ are given by \cite{Cardoso:2018ptl}
\bqn
f^{\epsilon_i}_t(r) &=& 1-\frac{2M}{r} + \epsilon_i \delta^i_1  \left(-\frac{1024M^9}{r^9} + \frac{1408M^{10}}{r^{10}}\right),\nb\\
f^{\epsilon_i}_r(r) &=& 1-\frac{2M}{r} + \epsilon_i \delta^i_1  \left(-\frac{4608M^9}{r^9}+\frac{8576M^{10}}{r^{10}}\right),\nb\\
\eqn
here the dimensionless coupling constant $\epsilon_i$ is defined via
\bqn
(\epsilon_1, \epsilon_2, \epsilon_3) = \left(\frac{1}{M^6 \Lambda^6}, \frac{1}{M^6 \tilde{\Lambda}^6}, \frac{1}{M^6 \Lambda_{-}^6}\right),
\eqn
with $M$ being the gravitational mass of the black hole spacetime. It is evident that the terms with $\tilde{C}$ in the gravitational action (\ref{action}) do not contribute to the spherical symmetric black hole spacetime. This is easy to understand, since the term $\tilde{C}$ is parity odd and it simply vanishes in the spherical symmetric background. 

The above spacetime describes a black hole with an event horizon located at 
\bqn
r_H=2M \left(1 + \frac{5 }{8}\epsilon_1 \right)+\mathcal{O}(\epsilon^2_i).
\eqn
As expected for higher-curvature modifications, these corrections decay rapidly with the radial distance $r$. The standard parameterized post-Newtonian parameters $\gamma, \beta$, etc., remain identical to those in general relativity. Likewise, GWs in this theory propagate at the speed of light when far from the source, and their dispersion relation coincides with that of GR \cite{Will:2014kxa, Berti:2009kk, LIGOScientific:2017bnn}.

The slowly rotating black holes in EFTGR are also considered in Ref.~\cite{Cardoso:2018ptl}, where the spin terms turn on the $\epsilon_2$ and $\epsilon_3$ corrections, which were absent for the static solutions. Also, the $\epsilon_3$ parameter adds an interesting “twist” to the solutions, making it the first purely gravitational example of a $\mathbb{Z}_2$-symmetry violating black hole solution. For simplicity, this paper focuses solely on the spherical symmetric case.

\section{Geodetic motion of a massive particle}\label{section3}
\renewcommand{\theequation}{3.\arabic{equation}} \setcounter{equation}{0}

In this section, we analyze the motion of a massive test particle in the spacetime of a static, spherically symmetric black hole within EFTGR. We start with the Lagrangian of a test particle,
\bqn
{\cal L }= \frac{1}{2}g_{\mu \nu} \frac{d x^\mu} {d \tau } \frac{d x^\nu}{d \tau},
\eqn
where $\tau$ denotes the affine parameter of the world line of the particle. For a massless particle, we have ${\cal L}=0$ and for a massive one ${\cal L} <0$. Then generalized momentum $p_\mu$ of the particle can be obtained via
\bqn
p_{\mu} = \frac{\partial {\cal L}}{\partial \dot x^{\mu}} = g_{\mu\nu} \dot x^\nu,
\eqn
which leads to four equations of motion for a particle with energy $E$ and angular momentum $L$,
\bqn
p_t &=& g_{tt} \dot t  = - E,\\
p_\phi &=& g_{\phi \phi} \dot \phi = L, \\
p_r &=& g_{rr} \dot r,\\
p_\theta &=& g_{\theta \theta} \dot \theta.
\eqn
Here, a dot denotes the derivative with respect to the affine parameter $\tau$ of the geodesics. From these expressions, we obtain
\bqn\lb{dot1}
\dot t = - \frac{ E  }{ g_{tt} } = \frac{E}{f^{\epsilon_i}_t(r)},\\
\lb{dot2}\dot \phi = \frac{ L}{g_{\phi\phi}} = \frac{L}{r^2 \sin^2\theta}.
\eqn
For timelike geodesics, substituting the above expressions of $\dot{t}$ and $\dot{\phi}$ into $g_{\mu\nu} x^\mu x^\nu =-1$ , we can get
\bqn
g_{rr} \dot r^2 + g_{\theta \theta} \dot \theta^2 &=& -1 - g_{tt} \dot t^2  - g_{\phi\phi} \dot \phi^2\nb\\
&=& -1 +\frac{E^2}{f^{\epsilon_i}_t(r)}- \frac{L^2}{r^2\sin^2\theta}.
\eqn

We are interested in the evolution of the particle in the equatorial plane, so we choose $\theta=\pi/2$ such that $\dot \theta=0$. Then the above expression can be simplified into the form of
\bqn\lb{rdot}
\dot r ^2 = \frac{f^{\epsilon_i}_r(r)}{f^{\epsilon_i}_t(r)}(E^2 - V_{\rm eff}(r)),
\eqn
where $V_{\rm eff}(r)$ denotes the effective potential and is given by
\bqn \lb{Veff}
V_{\rm eff}(r)= \left(1+\frac{L^2}{r^2}\right)f^{\epsilon_i}_t(r).
\eqn

One immediately observes that $V_{\rm eff}(r) \to 1$  as $r \to +\infty$, as expected for an asymptotically flat spacetime. Consequently, particles with $E>1$ can escape to infinity, while E=1 marks the critical energy separating bound from unbound orbits. In this sense, the maximum energy for bound orbits is $E=1$. Thus we can obtain the trajectory of a particle by integrating Eqs.~(\ref{dot1}),~(\ref{dot2}) and~(\ref{Veff}) to get $t$, $\phi$, and $r$ as functions of $\tau$. There is a convenient equation of motion that can be derived for numerical analysis from the $r-$ component of the geodesic equation, which is
\bqn
\ddot{r}= \frac{f^{\epsilon_i}_r(r)'}{2f^{\epsilon_i}_r(r)}\dot{r}^2-\frac{f^{\epsilon_i}_r(r) f^{\epsilon_i}_t(r)'E^2}{2f^{\epsilon_i}_t(r)^2}+\frac{f^{\epsilon_i}_r(r) L^2}{r^3}.
\eqn

This equation offers a convenient framework for computing orbital dynamics and GW emission. In Fig.~\ref{epsilon}, we plot the graphs of the effective potential as a function of the radial coordinate $r$ for $\epsilon_1=0.1$ and $\epsilon_1=0.05$. The curves from bottom to top explicitly show that the orbital angular momentum changes from $L_{\rm ISCO}$ to $L_{\rm MBO}$. The extremal points of $V_{\rm eff}$ are represented by the dashed line. It is shown that as the radial coordinate decreases, the location of the extremum of the effective potential shifts accordingly. When $d^2V_{\rm eff}/dr^2=0$, the potential admits a single extremum, which defines the radius of the innermost stable circular orbit (ISCO), with the associated orbital angular momentum denoted by $L_{\rm ISCO}$. For larger values of the orbital angular momentum, the effective potential develops two distinct extrema: an inner maximum corresponding to an unstable orbit and an outer minimum corresponding to a stable orbit.

\subsection{Marginally bound orbit}

Let us first consider a marginally bound orbit (MBO), which is defined as follows:
\bqn\lb{mbo}
V_{\rm eff}=E^2,\quad
\frac{dV_{\rm eff}}{dr}=0.
\eqn
By substituting Eq.~(\ref{Veff}) into Eq.~(\ref{mbo}), we can numerically solve the radius $r_{\rm MBO}$ and orbital angular momentum $L_{\rm MBO}$ for a fixed $\epsilon_1$.

\begin{figure}[htbp]
    \centering
    \includegraphics[width=0.9\linewidth]{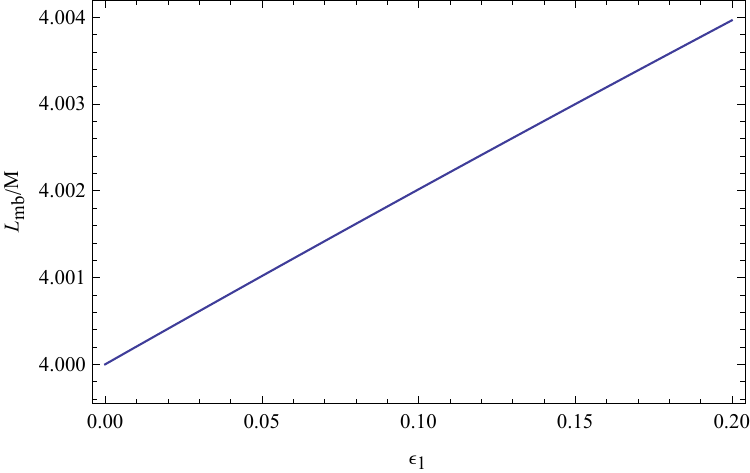}
    \includegraphics[width=0.9\linewidth]{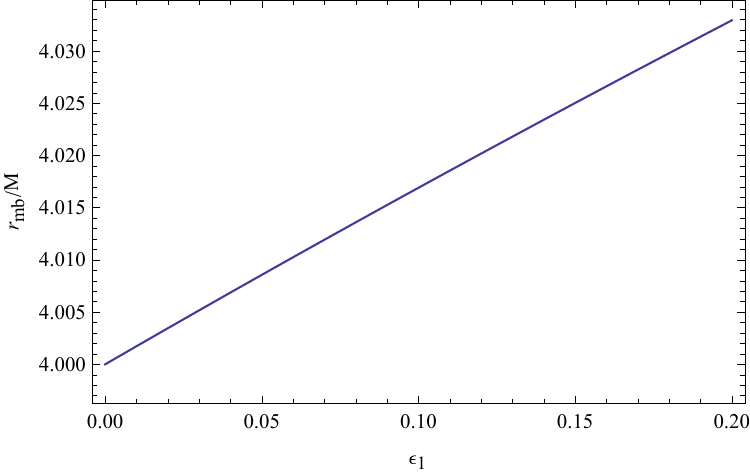}
    \caption{The angular momentum $L_{\text{MBO}}$ (upper panel) and
             radius $r_{\text{MBO}}$ (bottom panel), 
             for the marginally bound orbits.}
    \label{fig:MBO}
\end{figure}

As shown in Fig.~\ref{fig:MBO}, we plot the results of $L_{\rm MBO}$ and $r_{\rm MBO}$ with respect to the parameter $\epsilon_1$ of EFTGR black hole. It shows that the radius and angular momentum for the MBO increase with $\epsilon_1$. When $\epsilon_1=0$, these quantities reduce to those of a Schwarzschild black hole.

\subsection{Innermost stable circular orbit}

Let us turn to the ISCO, which is defined as
\bqn\lb{isco}
V_{\rm eff}=E^2, \quad
\frac{dV_{\rm eff}}{dr}=0,\quad
\frac{d^2V_{\rm eff}}{dr^2}=0.
\eqn
Since the metric is spherically symmetric, we can obtain the radius $r_{\rm ISCO}$ by solving Eq. (\ref{isco}).
Within the framework of EFTGR, these conditions for a black hole give rise to
\bqn
&&r_{\rm ISCO} = \frac{3f^{\epsilon_i}_t(r) \partial r f^{\epsilon_i}_t(r)^2}{2\partial r f^{\epsilon_i}_t(r)^2-f^{\epsilon_i}_t(r) \partial_r^2  f^{\epsilon_i}_t(r)},\\
&&E_{\rm ISCO} = \sqrt{\frac{2f^{\epsilon_i}_t(r)^2}{2f^{\epsilon_i}_t(r)-r \partial r f^{\epsilon_i}_t(r)}},\\
&&L_{\rm ISCO} = \sqrt{\frac{r^3   \partial r f^{\epsilon_i}_t(r)}{2f^{\epsilon_i}_t(r)-r \partial r f^{\epsilon_i}_t(r)}}.
\eqn

\begin{figure}[htbp]
    \centering
    \includegraphics[width=0.9\linewidth]{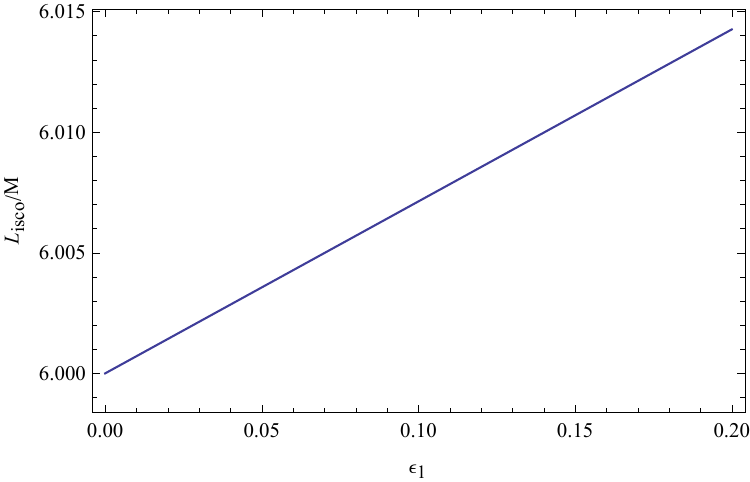}
    \includegraphics[width=0.9\linewidth]{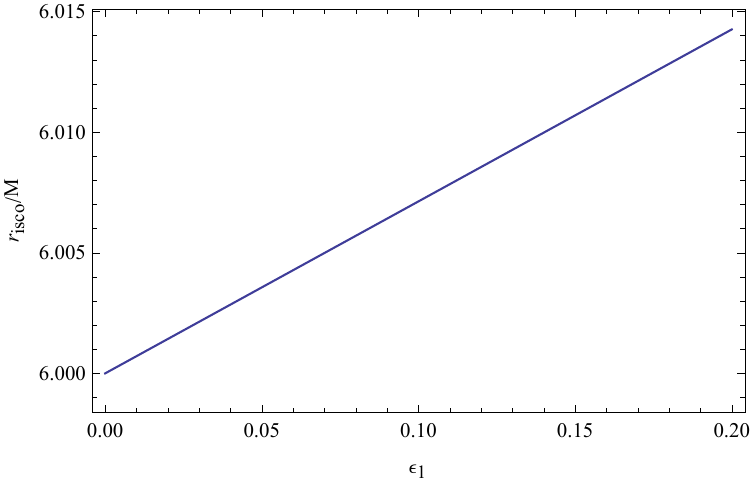}
    \includegraphics[width=0.9\linewidth]{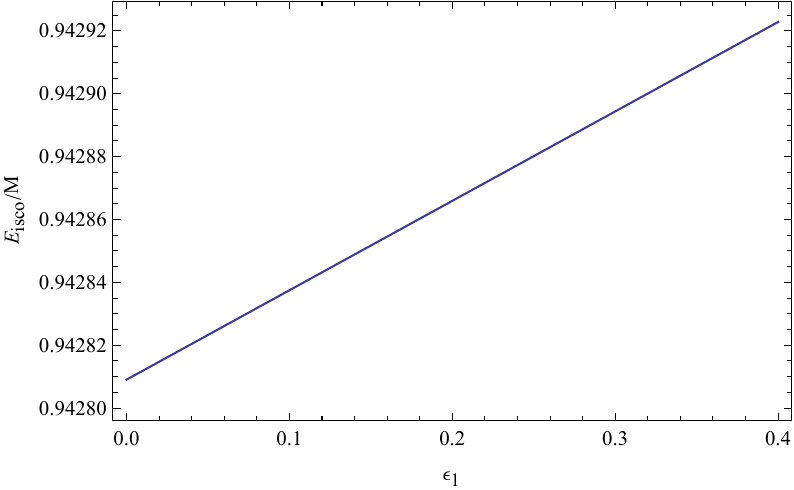}
    \caption{The angular momentum $L_{\rm ISCO}$ (upper panel), radius $r_{\rm ISCO}$ (middle panel), and energy $E_{\rm ISCO}$ (bottom panel) for the innermost stable circular orbits.}
    \label{fig:isco}
\end{figure}

In Fig.~\ref{fig:isco}, we plot the results of $L_{\rm ISCO}$, $r_{\rm ISCO}$, and $E_{\rm ISCO}$ with respect to the parameter $\epsilon_1$ of EFTGR black hole. It shows that the radius, angular momentum, and energy of the ISCO increase with $\epsilon_1$. When $\epsilon_1=0$, these quantities reduce to those of a Schwarzschild black hole.

\begin{figure*}[htbp]
    \centering
    \begin{minipage}{0.49\linewidth}
		\centering
		\includegraphics[width=0.9\linewidth]{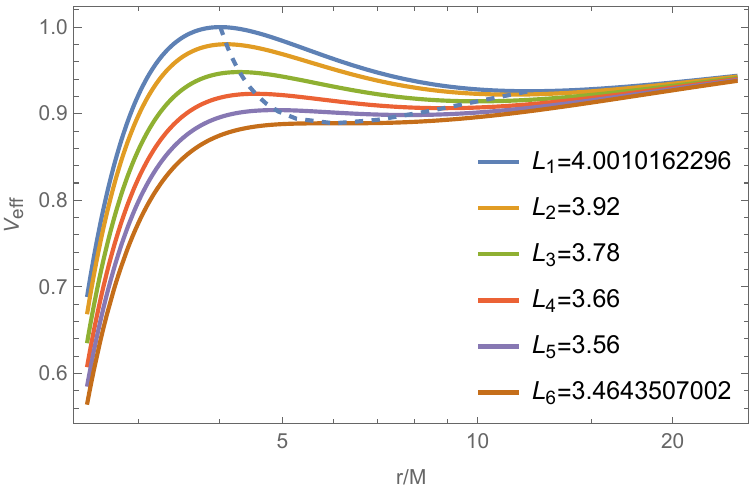}
		\label{epsilon0.05}
	\end{minipage}
    \begin{minipage}{0.49\linewidth}
		\centering
		\includegraphics[width=0.9\linewidth]{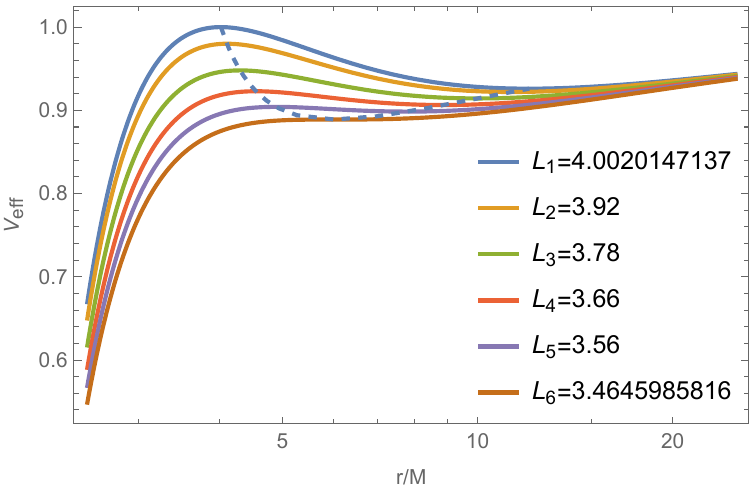}
		\label{epsilon0.1}
	\end{minipage}
    \caption{The effective potential $V_{\rm eff}$ as a function of $r/M$. The angular momentum $L/M$ varies from $L_{\rm ISCO}$ (bottom curve) to $L_{\rm MBO}$ (top curve). Dashed lines indicate the extremal points of the potential. Left and right panels correspond to EFTGR parameter values $\epsilon_1 = 0.01$ and $\epsilon_1 = 0.5$, respectively..}
    \label{epsilon}
\end{figure*}

\section{Periodic orbits}\label{section4}
\renewcommand{\theequation}{4.\arabic{equation}} \setcounter{equation}{0}
In this section, we discuss the periodic timelike orbits around EFTGR black holes. Following the taxonomy of Ref.~\cite{Levin:2008mq}, each periodic orbit is indexed by a triplet of integers $(z, w, v)$, denoting its zoom, whirl, and vertex characteristics. A periodic orbit arises when the ratio between the radial and azimuthal oscillation frequencies is rational, while generic orbits can be approximated by nearby periodic one \cite{Levin:2008ci}. This makes the study of periodic orbits a powerful tool for understanding the orbital structure and the associated GW emission. Within this framework, the frequency ratio is expressed as
\bqn\lb{qdefine}
q \equiv \frac{\omega_\phi}{\omega_r}-1 = w + \frac{v}{z},
\eqn
where $\omega_{r}$ and $\omega_{\phi}$ are the radial and azimuthal frequencies.
Since $\frac{\omega_\phi}{\omega_r}=\Delta \phi/(2\pi)$ with $\Delta \phi \equiv \oint d\phi$ denoting the azimuthal angle accumulated during one radial cycle. $q$ can be computed from the geodesic equations of EFTGR black holes as
\bqn\lb{q}
q &=& \frac{1}{\pi} \int_{r_1}^{r_2} \frac{\dot \phi}{\dot r} dr -1\nb\\
&=& \frac{1}{\pi} \int_{r_1}^{r_2} \frac{Ldr}{r^2\sqrt{\frac{f^{\epsilon_i}_r(r)}{f^{\epsilon_i}_t(r)}(E^2-V_{\rm eff})}}-1,
\eqn
with $r_1$ and $r_2$ being the radial turning points. By substituting (\ref{q}) into (\ref{qdefine}), it clearly shows that the factors influencing the rational number of periodic orbits are the particle's energy $E$, its orbital angular momentum $L$ and the $\epsilon_1$.

\begin{figure*}[htbp]
\centering
{
\begin{minipage}[b]{.3\linewidth}
\centering
\includegraphics[scale=0.3]{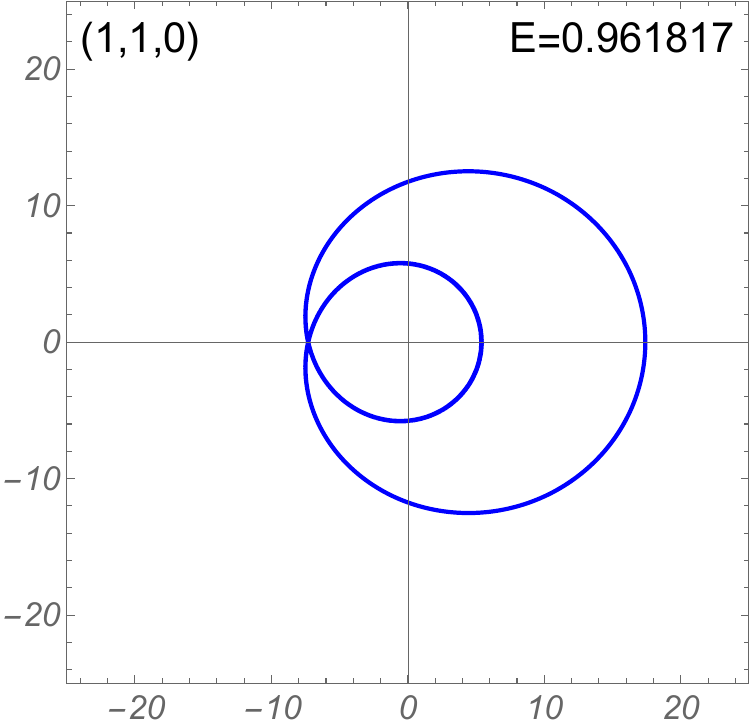}
\end{minipage}
}
{
\begin{minipage}[b]{.3\linewidth}
\centering
\includegraphics[scale=0.3]{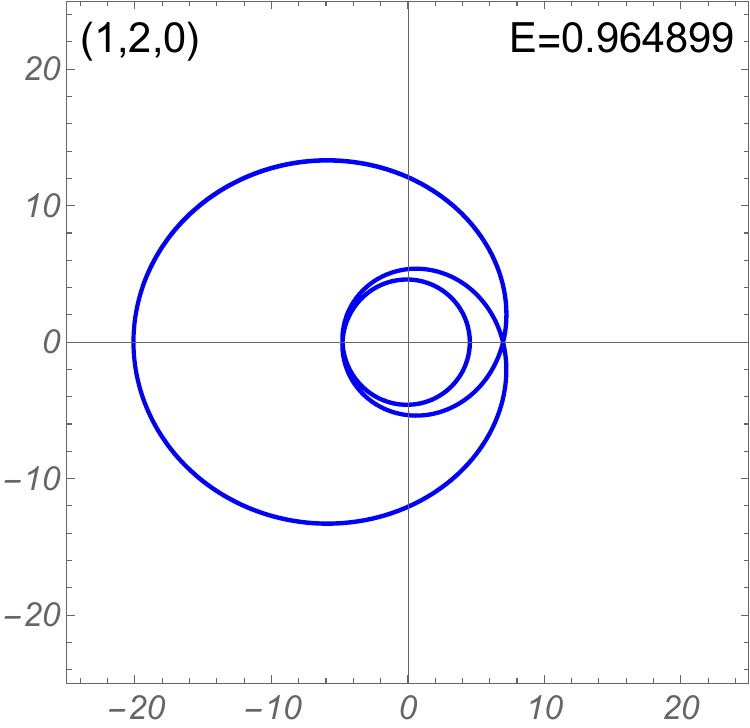}
\end{minipage}
}
{
\begin{minipage}[b]{.3\linewidth}
\centering
\includegraphics[scale=0.3]{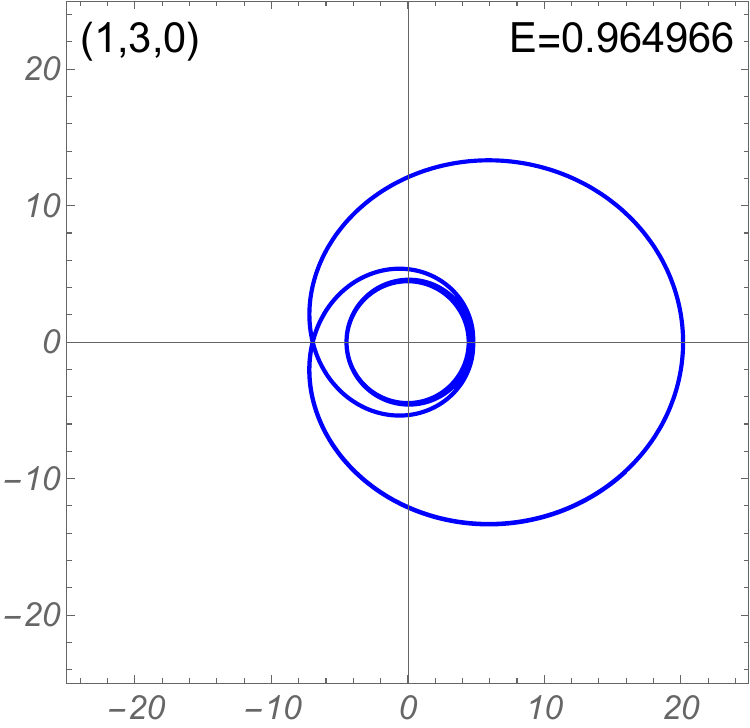}
\end{minipage}
}
{
\begin{minipage}[b]{.3\linewidth}
\centering
\includegraphics[scale=0.3]{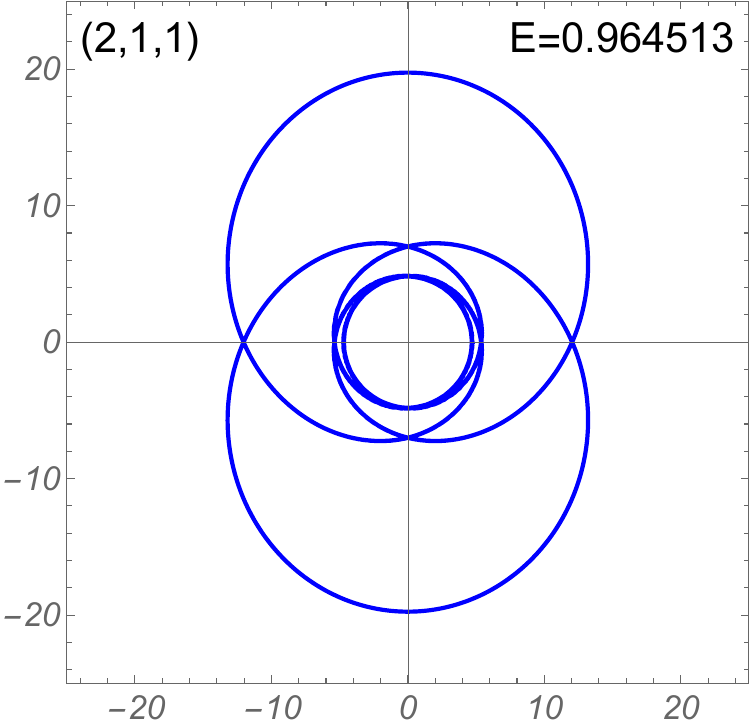}
\end{minipage}
}
{
\begin{minipage}[b]{.3\linewidth}
\centering
\includegraphics[scale=0.3]{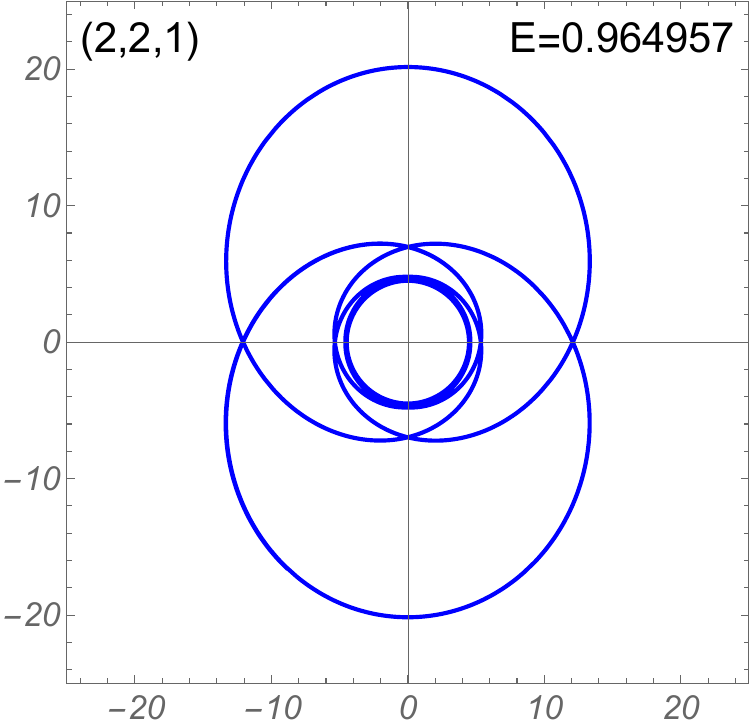}
\end{minipage}
}
{
\begin{minipage}[b]{.3\linewidth}
\centering
\includegraphics[scale=0.3]{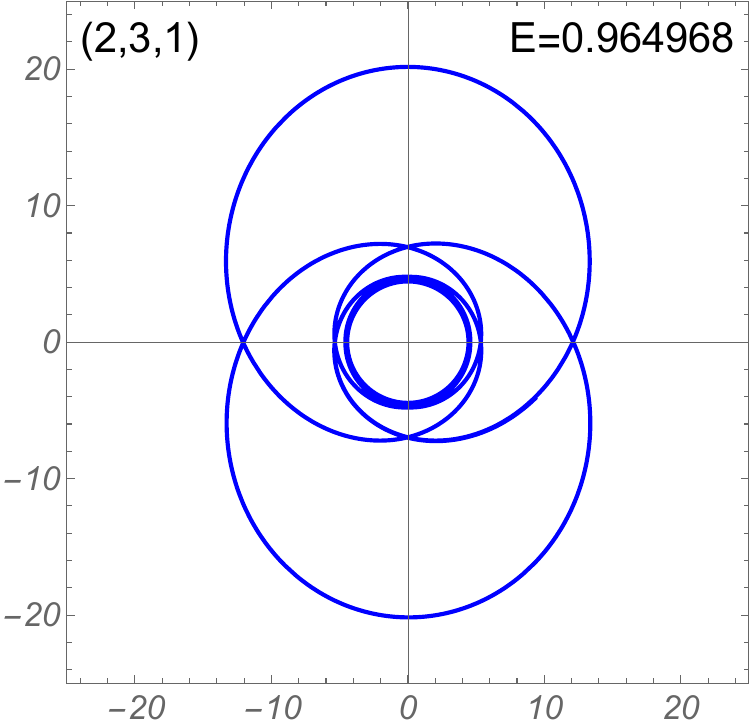}
\end{minipage}
}
{
\begin{minipage}[b]{.3\linewidth}
\centering
\includegraphics[scale=0.3]{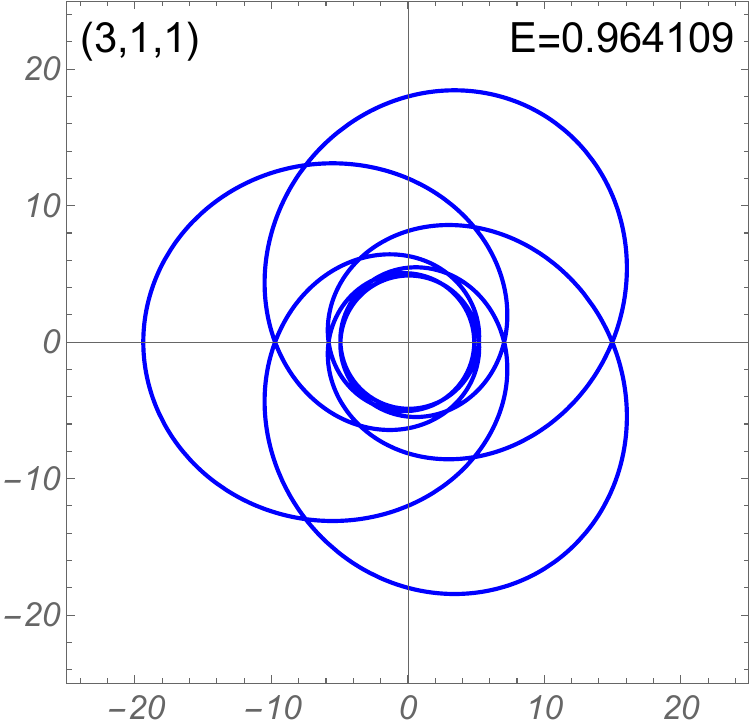}
\end{minipage}
}
{
\begin{minipage}[b]{.3\linewidth}
\centering
\includegraphics[scale=0.3]{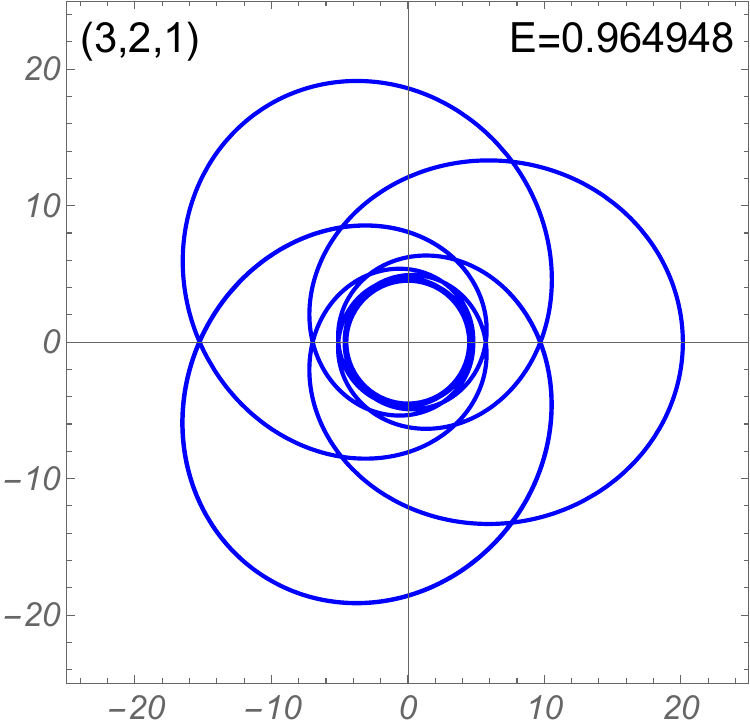}
\end{minipage}
}
{
\begin{minipage}[b]{.3\linewidth}
\centering
\includegraphics[scale=0.3]{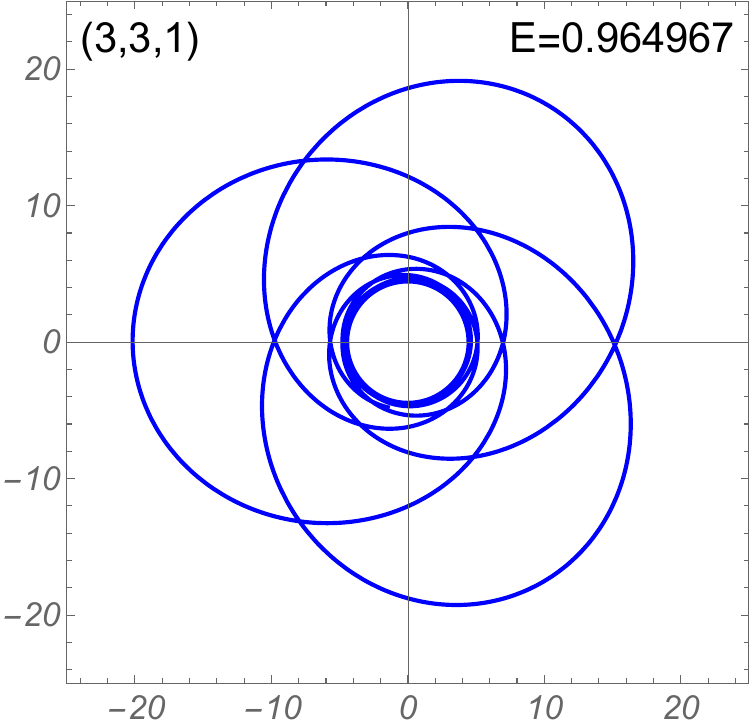}
\end{minipage}
}
{
\begin{minipage}[b]{.3\linewidth}
\centering
\includegraphics[scale=0.3]{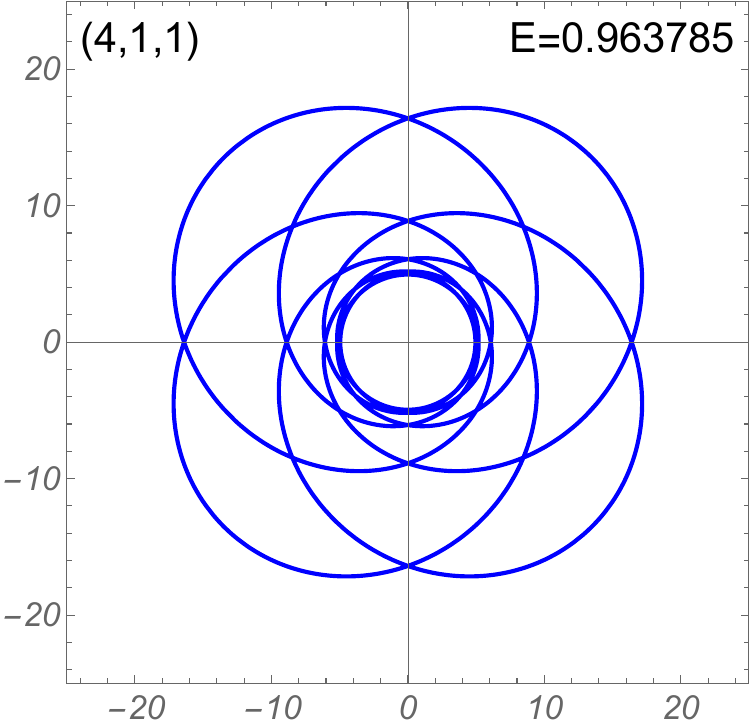}
\end{minipage}
}
{
\begin{minipage}[b]{.3\linewidth}
\centering
\includegraphics[scale=0.3]{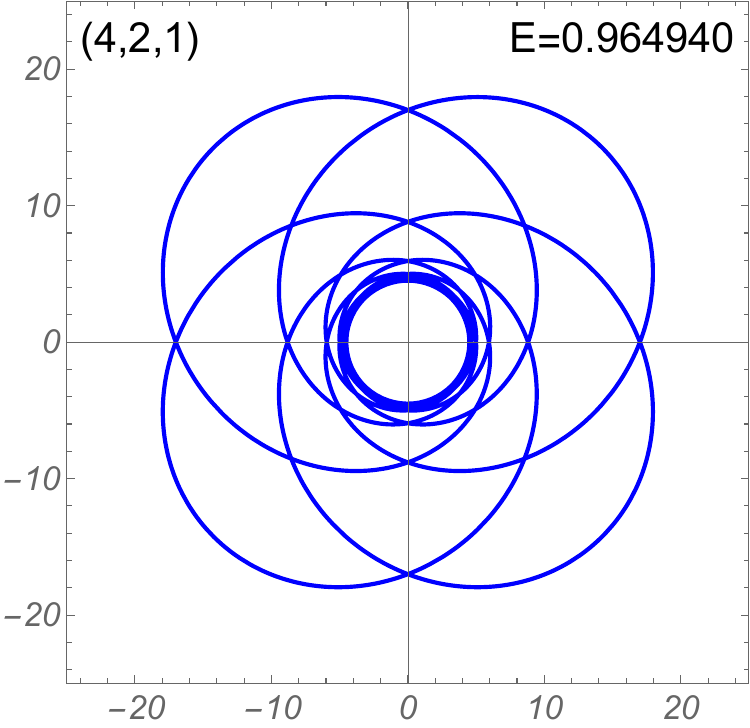}
\end{minipage}
}
{
\begin{minipage}[b]{.3\linewidth}
\centering
\includegraphics[scale=0.3]{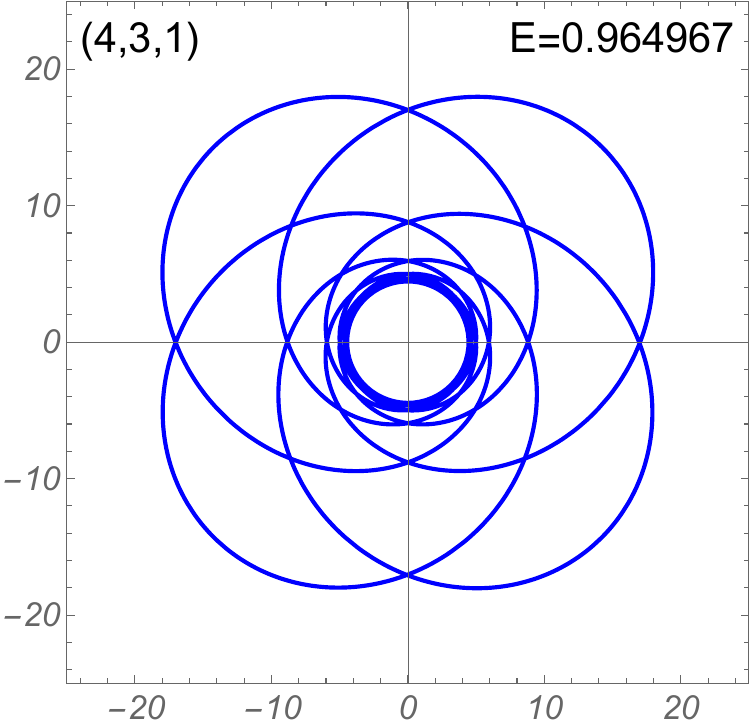}
\end{minipage}
}
\caption{Periodic orbits around a black hole in EFTGR for various combinations of ($z,w,v$). Each row corresponds to a fixed value of $z$, increasing from top to bottom, while each column corresponds to a fixed value of $w$, increasing from left to right. The system is evaluated at energy $E=0.96$ with EFTGR parameter $\epsilon_{1}=0.01$. Orbits are projected onto the equatorial plane $(\theta  = \pi/2)$ using Cartesian coordinates $(x, y) = (r \sin \phi, r \cos \phi)$.}
\label{epsilon0.01}
\end{figure*}

\begin{figure*}[htbp]
\centering
{
\begin{minipage}[b]{.3\linewidth}
\centering
\includegraphics[scale=0.3]{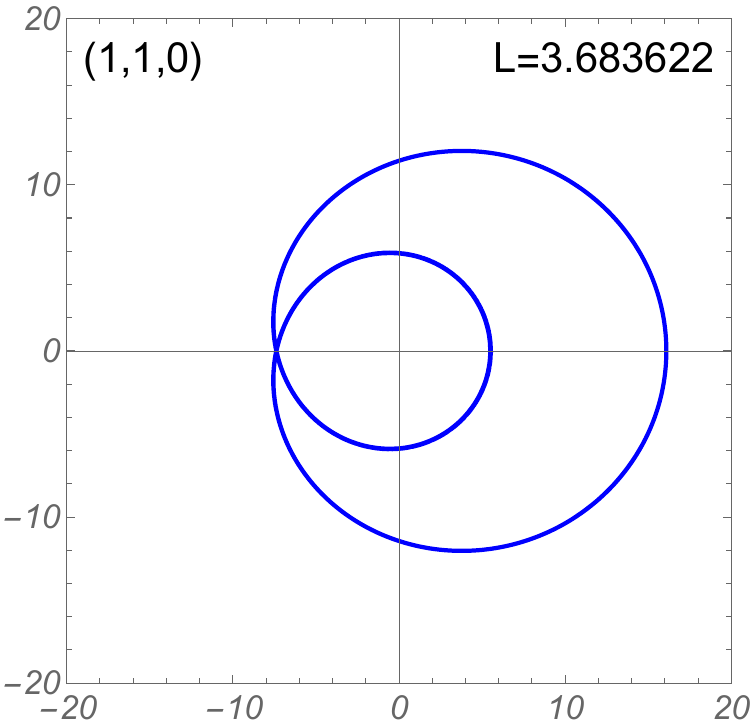}
\end{minipage}
}
{
\begin{minipage}[b]{.3\linewidth}
\centering
\includegraphics[scale=0.3]{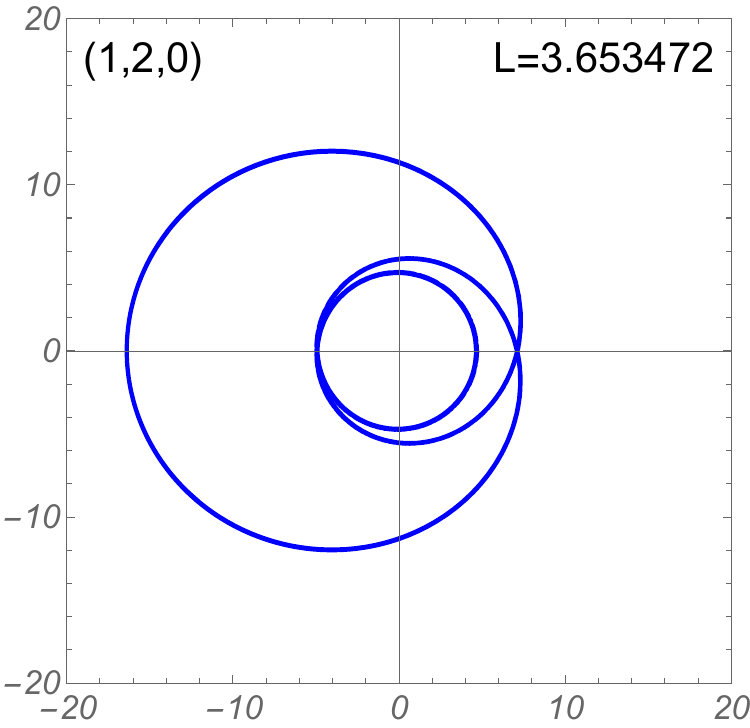}
\end{minipage}
}
{
\begin{minipage}[b]{.3\linewidth}
\centering
\includegraphics[scale=0.3]{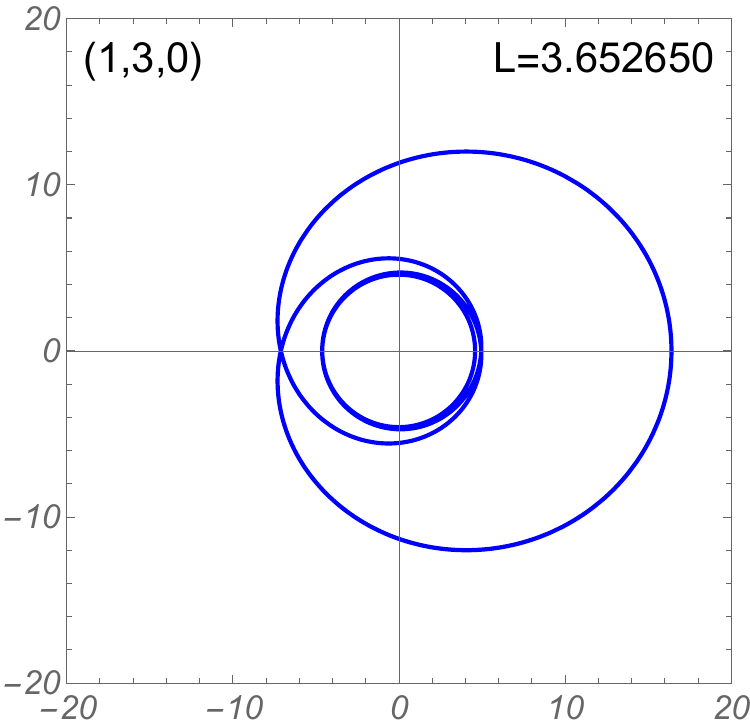}
\end{minipage}
}
{
\begin{minipage}[b]{.3\linewidth}
\centering
\includegraphics[scale=0.3]{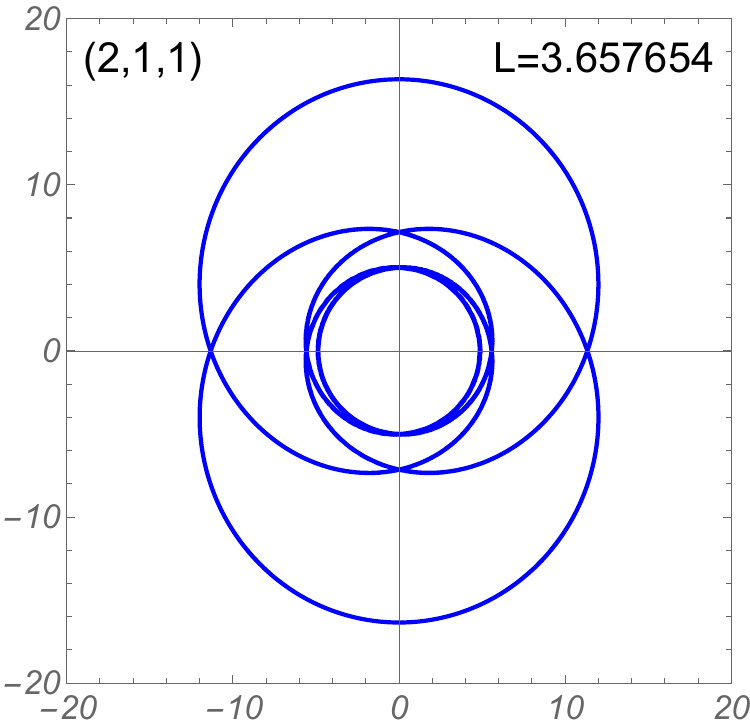}
\end{minipage}
}
{
\begin{minipage}[b]{.3\linewidth}
\centering
\includegraphics[scale=0.3]{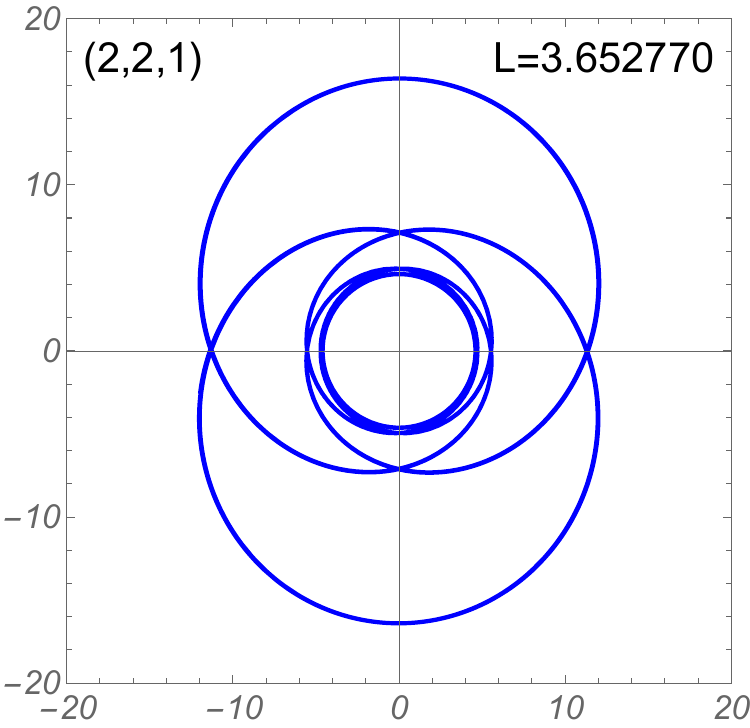}
\end{minipage}
}
{
\begin{minipage}[b]{.3\linewidth}
\centering
\includegraphics[scale=0.3]{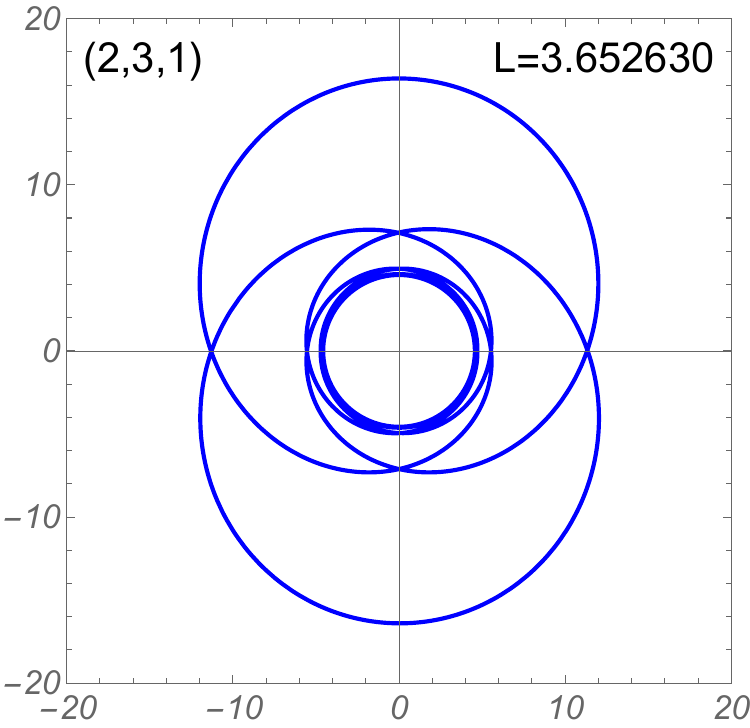}
\end{minipage}
}
{
\begin{minipage}[b]{.3\linewidth}
\centering
\includegraphics[scale=0.3]{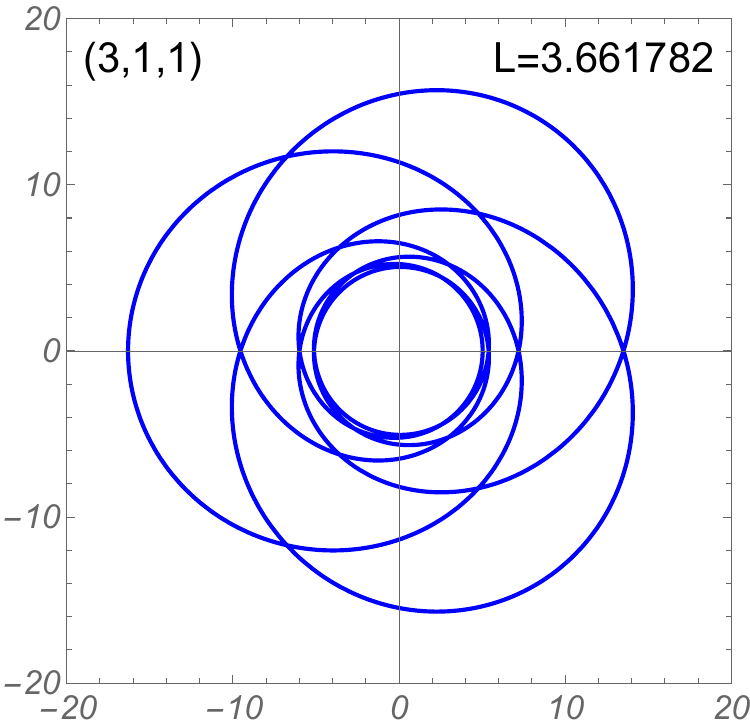}
\end{minipage}
}
{
\begin{minipage}[b]{.3\linewidth}
\centering
\includegraphics[scale=0.3]{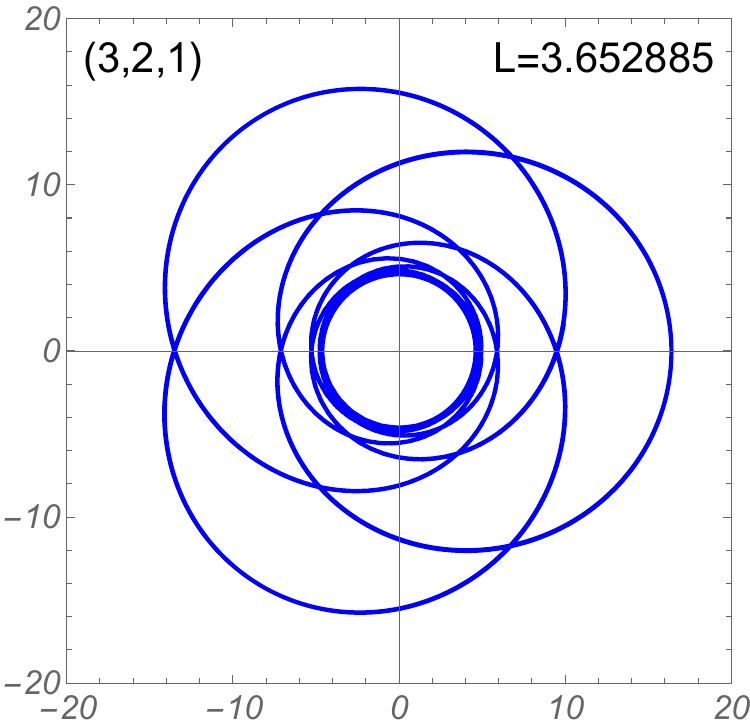}
\end{minipage}
}
{
\begin{minipage}[b]{.3\linewidth}
\centering
\includegraphics[scale=0.3]{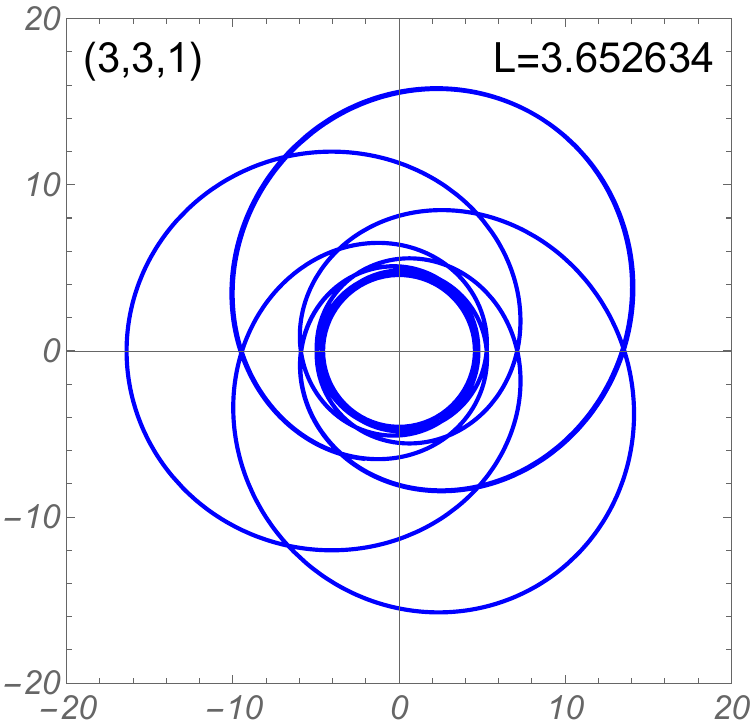}
\end{minipage}
}
{
\begin{minipage}[b]{.3\linewidth}
\centering
\includegraphics[scale=0.3]{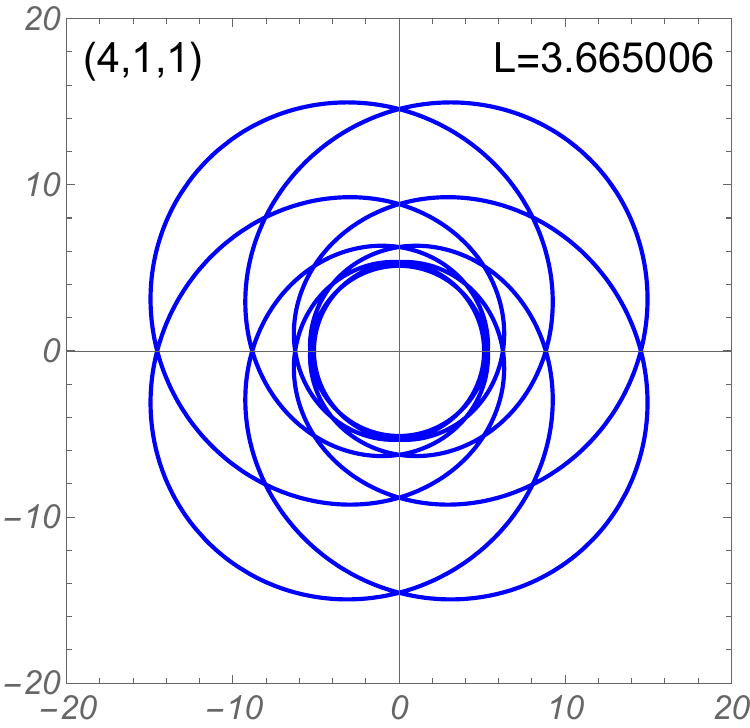}
\end{minipage}
}
{
\begin{minipage}[b]{.3\linewidth}
\centering
\includegraphics[scale=0.3]{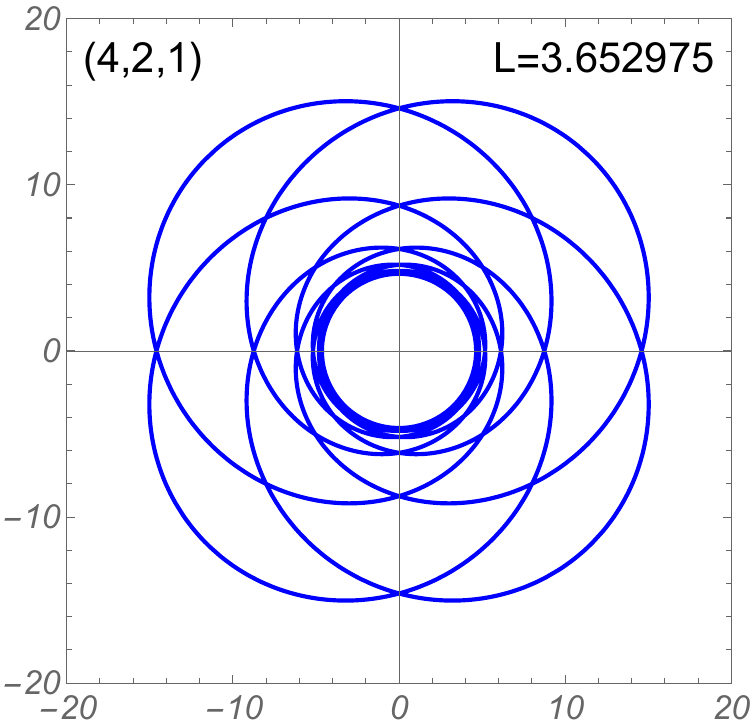}
\end{minipage}
}
{
\begin{minipage}[b]{.3\linewidth}
\centering
\includegraphics[scale=0.3]{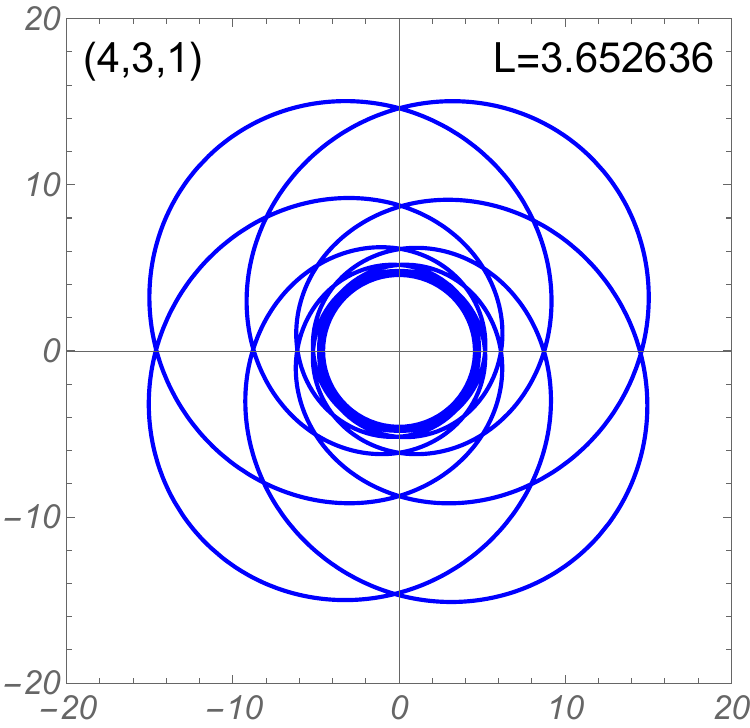}
\end{minipage}
}
\caption{Periodic orbits around a black hole in EFTGR for various combinations of ($z,w,v$). Each row corresponds to a fixed value of $z$, increasing from top to bottom, while each column corresponds to a fixed value of $w$, increasing from left to right. The system is evaluated at angular momentum $L=3.7$ with EFTGR parameter $\epsilon_{1}=0.01$. Orbits are projected onto the equatorial plane $(\theta  = \pi/2)$ using Cartesian coordinates $(x, y) = (r \sin \phi, r \cos \phi)$.}
\label{epsilon0.01L}
\end{figure*}

 In Figs.~\ref{epsilon0.01} and~\ref{epsilon0.01L}, we illustrate the periodic orbits using different combinations of integers $(z, w, v)$ for the EFTGR black hole, respectively. The integer $z$ governs the number of leaves in the orbital trajectory, $w$ determines the number of small loops within the orbit, and $v$ dictates the orientation (clockwise or counterclockwise) of the system. Higher values of $z$ result in larger blade profiles and more complex trajectories. In Fig.~\ref{epsilon0.01}, we show periodic orbits for fixed energy $E=0.96$ and $\epsilon_1=0.01$. In Fig.~\ref{epsilon0.01L}, we show periodic orbits for fixed angular momentum $L=3.7$ and $\epsilon_1=0.01$. Both figures are organized such that the parameter $z$ remains constant along each row and increases sequentially from the top row to the bottom row. In contrast, the parameter $w$ remains constant along each column and increases from left to right. We can see that in both figures the trajectory of the orbits shows exactly periodic behavior, which is determined by the given combinations of the three integers $(z,w,v)$.

\section{Gravitational Radiation in effective field theory extension of general relativity}\label{section5}
\renewcommand{\theequation}{5.\arabic{equation}} \setcounter{equation}{0}

In this section, we examine the gravitational radiation emitted by the periodic orbits of a test particle orbiting a supermassive black hole described within EFTGR. As the particle approaches  the central supermassive black hole, it experiences a progressively stronger gravitational field, causing changes in the curvature of spacetime and generating GWs. The loss of energy and angular momentum via gravitational radiation drives the particle into a slow inspiral toward the black hole. During a few orbital cycles, the energy and angular momentum lost by the smaller mass during its motion are negligible compared to the total budget, which motivates the adoption of the adiabatic approximation~\cite{Grossman:2011im, Zi:2023qfk, Zi:2024dpi, Hughes:2005qb, Sundararajan:2007jg, Hughes:1999bq}.  The adiabatic approximation assumes that during a single orbital period, the system’s energy and angular momentum remain effectively constant, so the influence of gravitational radiation on the motion of the smaller body can be neglected. Within this framework, the trajectory of the test particle is governed by the geodesic equations, and the corresponding gravitational waveforms are obtained using the quadrupole formula ~\cite{Babak:2006uv}. The procedure involves two steps: first, the geodesic equations in the EFTGR background are numerically solved to obtain the trajectory of the particle; second, this trajectory is inserted into the quadrupole expression to compute the waveform emitted. This method provides a first step toward analyzing EMRI gravitational waves and their ability to probe orbital behavior and central black hole properties.
 
As the inspiral progresses toward the horizon, the particle undergoes increasingly rapid oscillations that imprint distinctive structures on the gravitational waveform. These signals encode valuable information about the near-horizon regime and serve as sensitive probes of black hole properties. To quantify these features, we now turn to the quadrupole formalism, in which the metric perturbation $h_{ij}$ is related to the symmetric and trace-free mass quadrupole moment as
\bqn
h_{ij}=\frac{1}{A}\frac{d^2I_{ij}}{dt^2},
\eqn
where $A=c^4D_L/(2G)$, with $D_L$ being the luminosity distance from the EMRI system to the detector and $c$ the speed of light. By numerically solving the geodesic equations of motion, the trajectory $Z_i(t)$ of the smaller object in the curved spacetime can be determined, which serves as the essential input for constructing the corresponding gravitational waveform. For a smaller object with mass $m$ following a trajectory $Z^i(t)$, the $I_{ij}$ is given by 
\bqn
I^{ij}=m\int d^3x^ix^j\delta^3(x^i-Z^i(t)).
\eqn

The choice of coordinate system plays a crucial role in the calculation and interpretation of gravitational waveforms. While the geodesic equations are often solved in Boyer-Lindquist coordinates $(r, \theta, \phi)$, the waveform is typically expressed in a detector-adapted coordinate system $(X,Y,Z)$. The transformation from 	Boyer-Lindquist coordinates to Cartesian coordinates is given by~\cite{Babak:2006uv}
\bqn
x=r \sin\theta \cos\phi,\ y=r \sin\theta \sin\phi, \ z=r\cos\theta.
\eqn
This allows us to project the trajectory of the small object onto a Cartesian grid. The metric perturbations $h_{ij}$, representing the GWs, are then calculated using the second time derivative of the mass quadrupole moment $I_{ij}$ as
\bqn
h_{ij}=\frac{2m}{D_L}\frac{d^2I_{ij}}{dt^2}=\frac{2m}{D_L}(a_ix_j+a_jx_i+2v_iv_j),
\eqn
where $v_i$ and $a_i$ represent the spatial velocity and acceleration of the small object, respectively.

The polarizations of the waveforms, which are usually denoted as $h_+$ and $h_\times$ to distinguish these two different modes, can be written as the combination of $h_{ij}$ via the transformation of coordinates from $(x,y,z)$ to $(X,Y,Z)$. As a result, the corresponding plus $h_{+}$ and cross $h_{\times}$ polarizations of the GW are given by
\bqn
h_+&=&\left(h_{\Theta\Theta}-h_{\Phi\Phi}\right)/2,\\
h_\times &=& h_{\Theta\Phi},
\eqn
where the components $h_{\Theta\Theta}$, $h_{\Phi\Theta}$ and $h_{\Phi\Phi}$ are given by
\bqn
h_{\Theta\Theta}&=& \cos^2 \Theta \left[h_{xx} \cos^2\Phi + h_{xy} \sin2\Phi + h_{yy} \sin^2\Phi \right]\nb\\
&& + h_{zz} \sin^2\Theta - \sin2\Theta \left[h_{xz}\cos\Phi+h_{yz}\sin\Phi \right],\nb\\
h_{\Phi\Theta}&=&\cos\Theta \left[\frac{1}{2}h_{xx}\sin2\Phi+h_{xy}\cos2\Phi+\frac{1}{2}h_{yy}\sin2\Phi \right]\nb\\
&&+\sin\Theta \left[h_{xz}\sin\Phi-h_{yz}\cos\Phi \right],\nb\\
h_{\Phi\Phi}&=&h_{xx}\sin^2\Phi-h_{xy}\sin2\Phi+h_{yy}\cos^2\Phi,
\eqn
where $\Theta$ is the inclination angle and $\Phi$ is the latitude angle.

To illustrate the GW waveforms of different periodic orbits and how the $\epsilon_1$ can affect it, we consider an EMRI system consisting of a small object with mass $m=10 M_\odot$ and a supermassive black hole with mass $M=10^{7} M_\odot$, where $M_\odot$ is the solar mass. For simplicity, we set the inclination angle $\Theta=\pi/4$ and the latitude angle $\Phi=\pi/4$. The luminosity distance is set to $D_{\rm L}=200\;{\rm Mpc}$.

\begin{figure*}[htbp]
    \centering
    \begin{minipage}{0.49\linewidth}
		\centering
		\includegraphics[width=0.75\linewidth]{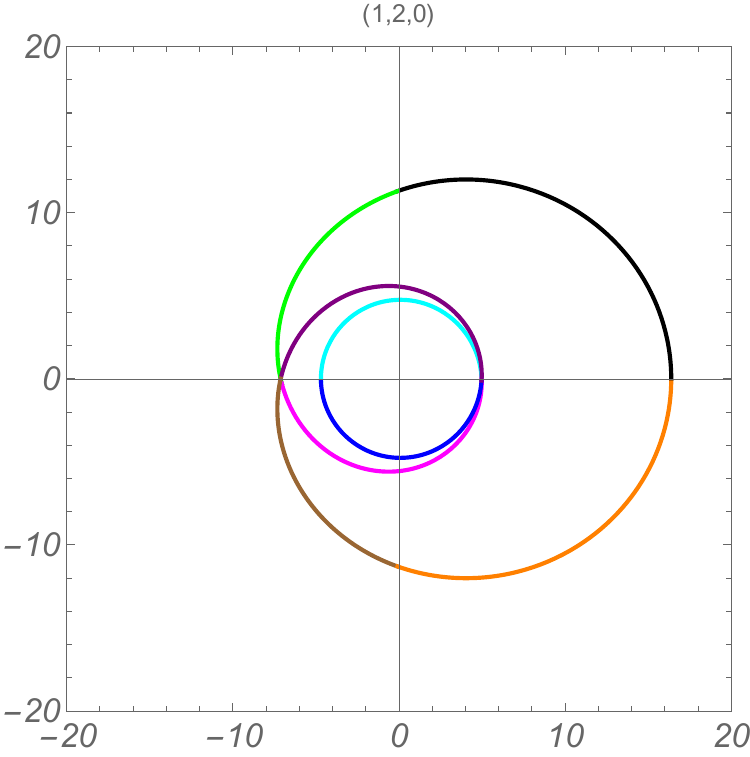}\hspace{15mm}
		\label{orbit120}
	\end{minipage}
    \begin{minipage}{0.49\linewidth}\vspace{8mm}\hspace{-15mm}
		\centering
		\includegraphics[width=0.9\linewidth]{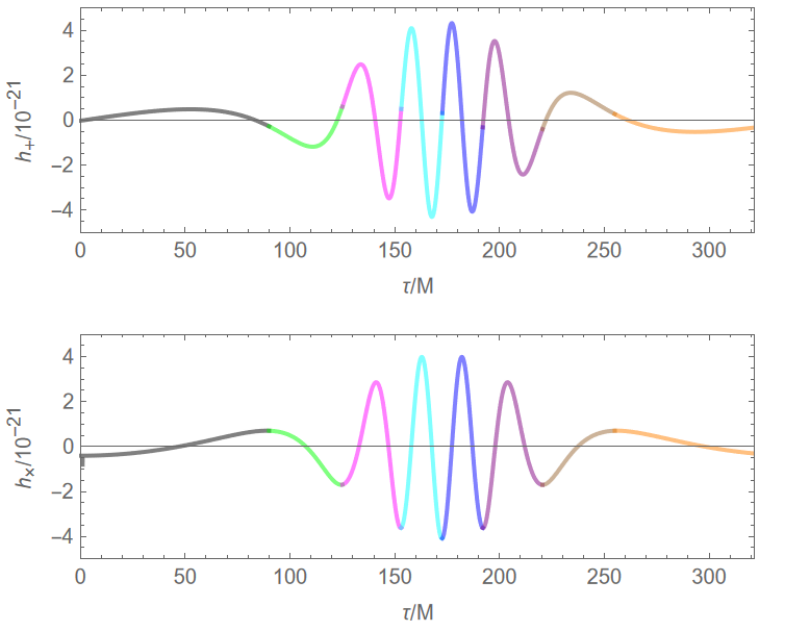}
		\label{120right}
	\end{minipage}
    \caption{The left figure shows a particle traveling from an apastron to another in a typical periodic orbit around an EFTGR black hole with ($z, w, v$) = (1, 2, 0); The right figures represent the $h_{+}$ and $h_{\times}$ modes of the GW of the black hole in EFTGR, with $q=2$. In all these figures, different colors represent different periods of the orbit. In the left panel, orbit is projected onto the equatorial plane $(\theta  = \pi/2)$ using Cartesian coordinates $(x, y) = (r \sin \phi, r \cos \phi)$.}
    \label{120waveform}
\end{figure*}

\begin{figure*}[htbp]
    \centering
    \begin{minipage}{0.49\linewidth}
		\centering
		\includegraphics[width=0.75\linewidth]{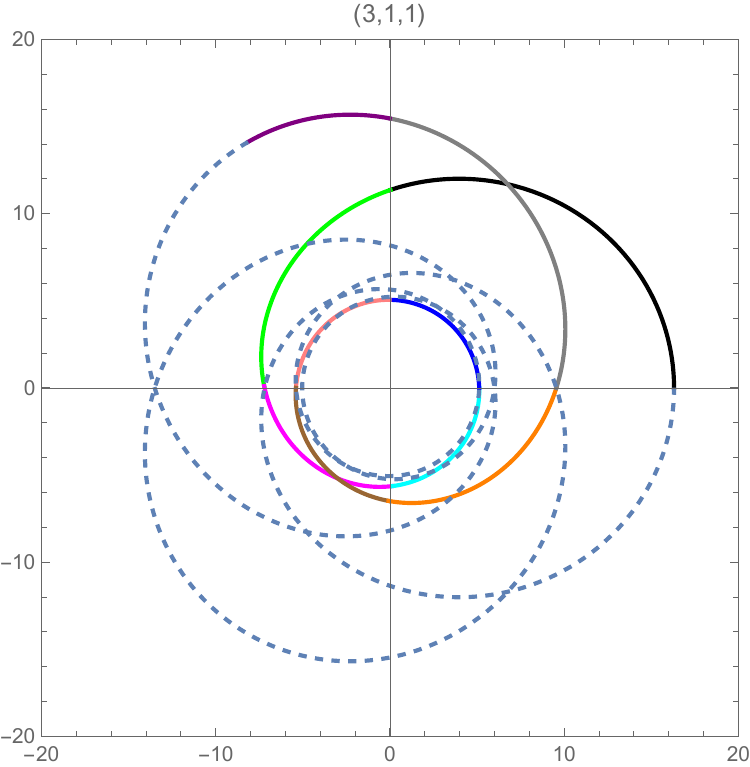}\hspace{15mm}
		\label{orbit311}
	\end{minipage}
    \begin{minipage}{0.49\linewidth}\vspace{8mm}\hspace{-15mm}
		\centering
		\includegraphics[width=0.9\linewidth]{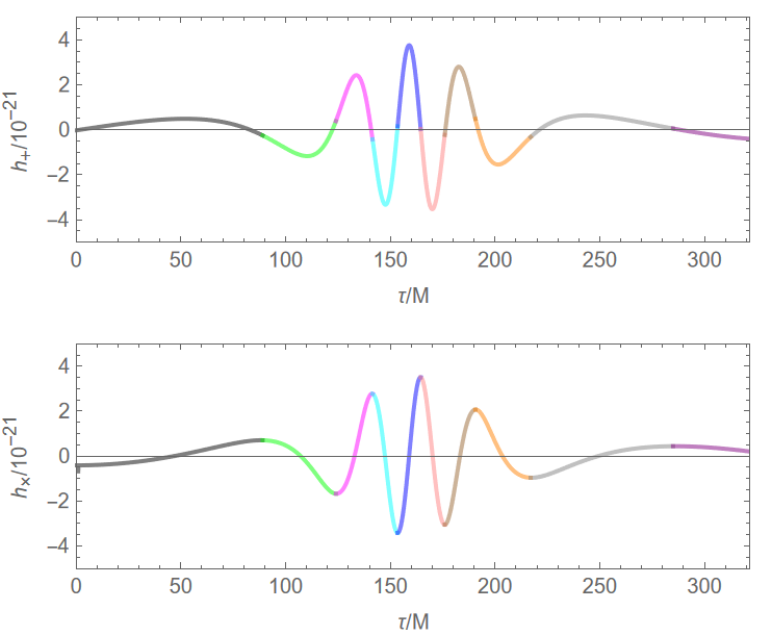}
		\label{311right}
	\end{minipage}
    \caption{The left figure shows a particle traveling from an apastron to another in a typical periodic orbit around an EFTGR black hole with ($z, w, v$) = (3, 1, 1); The right figures represent the $h_{+}$ and $h_{\times}$ modes of the GW of the black hole in EFTGR, with $q=\frac{4}{3}$. In all these figures, different colors represent different periods of the orbit. In the left panel, orbit is projected onto the equatorial plane $(\theta  = \pi/2)$ using Cartesian coordinates $(x, y) = (r \sin \phi, r \cos \phi)$.}
    \label{311waveform}
\end{figure*}

In Figs.~\ref{120waveform} and~\ref{311waveform}, we show waveforms emitted by a periodic orbit in EFTGR black hole, with $(z,w,v) = (1,2,0)$ and $(z,w,v) = (3,1,1)$. In both figures, we utilize various colors to emphasize the connection between the gravitational waveforms and the periodic orbits, facilitating a comprehensive examination of how the periodic orbits influence the resulting gravitational radiation. We can see that the GW waveform shows zoom-whirl behaviors in an entire circle. By comparing the figures of the periodic orbits with the figures of the GW, we can see that the amplitude of the GW increases as the object travels into the perihelion and decreases as it moves farther from the perihelion. In the beginning stage of the trajectory, the gravitational field is weak as the small object is far from the black hole, thus the waveforms are smooth. When the small object travels into the strong field regime, the trajectory gets distorted, with an evident change in both the frequency and amplitude of the GW. In the final stage, the small object is near the event horizon, and the GW signal reaches its peak with a sharp burst in both frequency and amplitude. Consequently, the periodic orbits and their corresponding GW signals carry significant information about the EMRI system, which can help understand the physical properties of black holes and their surrounding areas. 

\begin{figure*}[htbp]
    \centering
    \begin{minipage}{0.49\linewidth}
		\centering
		\includegraphics[width=0.75\linewidth]{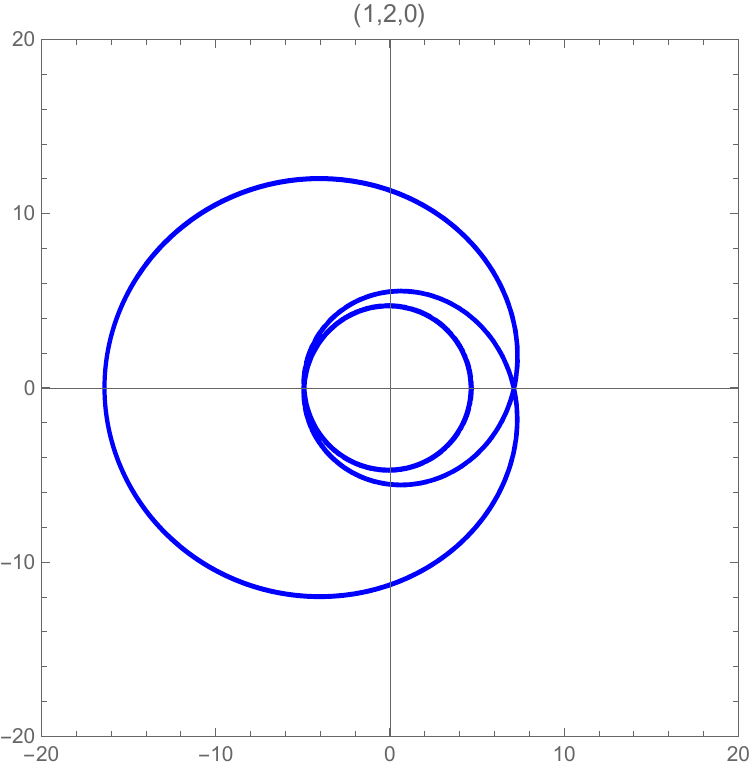}\hspace{15mm}
	\end{minipage}
    \begin{minipage}{0.49\linewidth}
		\centering
        \includegraphics[width=9cm,height=6cm]{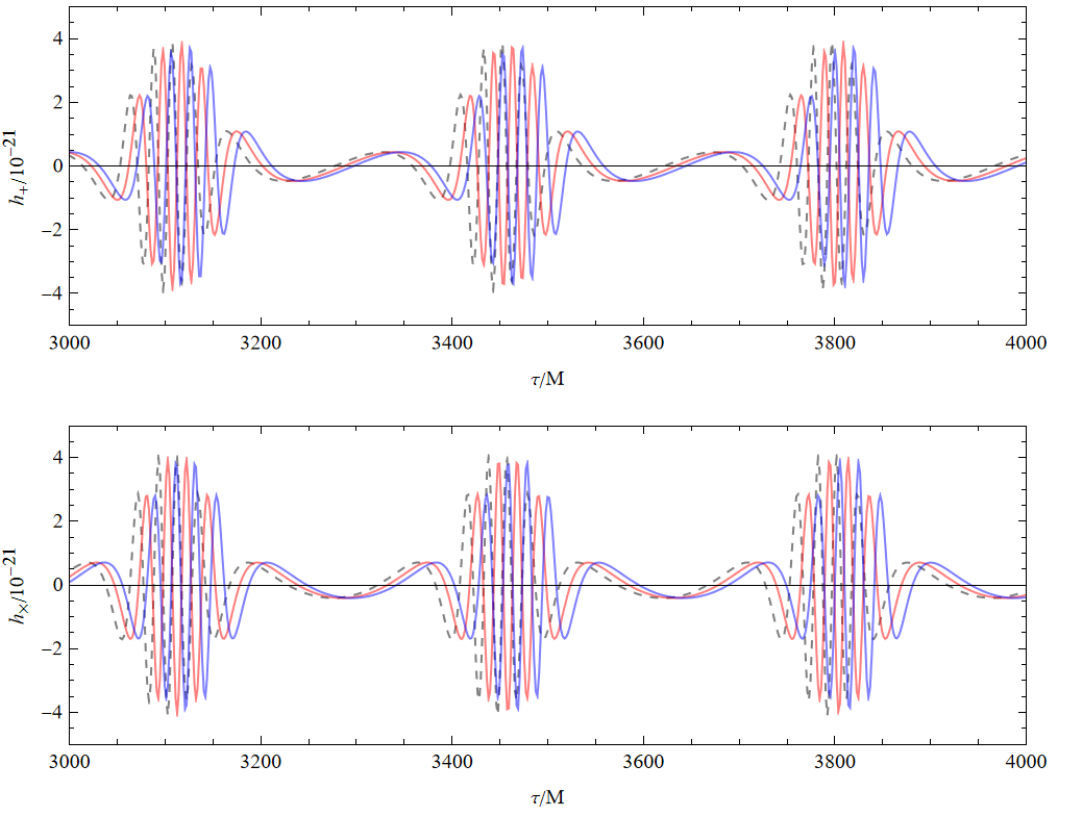}\vspace{-8mm}\hspace{-5mm}
	\end{minipage}
    \caption{The left figure is a figure showing a typical orbit around an EFTGR black hole with ($z, w, v$) = (1, 2, 0); The right figure represents the $h_{+}$ and $h_{\times}$ modes of GW of a black hole in EFTGR, with $q=2$. The dashed line corresponds to the GW signal of a Schwarzschild black hole ($\epsilon_1=0$), while the orange and blue lines represent the GW signals of EFTGR black holes with $\epsilon_1=0.5$ and $\epsilon_1=1$, respectively. To more clearly illustrate the effect of $\epsilon_1$ on the GW waveform generated by periodic orbits, we set the $\tau/M$ to $3000–4000$. In the left panel, orbit is projected onto the equatorial plane $(\theta  = \pi/2)$ using Cartesian coordinates $(x, y) = (r \sin \phi, r \cos \phi)$.}
    \label{120GWdifferenceE}
\end{figure*}

In Figs.~\ref{120GWdifferenceE}, we plot a typical periodic orbit with $(z,w,v)=(1,2,0)$ and their $h_+$ and $h_\times$ modes with different values of $\epsilon_1$. First, a clear relationship emerges between waveform and orbit in which the quiet phases correspond to ``leaves", while intense oscillations align with whirl-dominated segments of the trajectory. Secondly, the effect of $\epsilon_1$ is weak, as changes in phase become obvious after many periods. The parameter $\epsilon_1$ mainly affects the phase of the GW signals and offers no change in amplitudes. These findings offer potential for detecting and distinguishing physical signatures in future detections.

\section{Discussions and Conclusions}\label{section6}
\renewcommand{\theequation}{6.\arabic{equation}} \setcounter{equation}{0}

In this paper, we have investigated the periodic orbits and their associated waveforms in EFTGR. We began by deriving the geodesic equations for particles orbiting a black hole in EFTGR, which deviate from those in Schwarzschild spacetime. Through a numerical analysis of the effective potential, which is shown in Fig.~\ref{epsilon}, we characterized the properties of MBOs and ISCOs. As shown in Figs.~\ref{fig:MBO} and ~\ref{fig:isco}, both the radius and angular momentum of the MBO and the ISCO increase with the coupling parameter $\epsilon_1$. In the limit $\epsilon_1=0$, these quantities reduce to their Schwarzschild counterparts. 

Building on the behavior of MBOs and ISCOs, we further explore periodic orbits in EFTGR black hole spacetimes. To classify these orbits, We adopted the taxonomy introduced in Ref.~\cite{Levin:2008mq}, where each periodic orbit is labled by a triplet $(z,w,v)$. Figures~\ref{epsilon0.01} and \ref{epsilon0.01L} present representative periodic trajectories for a black hole in EFTGR with $\epsilon_1=0.01$, for fixed values of $E$ and $L$.

We then examined the GW emission from these periodic orbits. Figs.~\ref{120waveform} and~\ref{311waveform} use color-coded segments to illustrate the correspondence between orbital phases and waveform features, providing information on how orbital dynamics shape GWs. In both cases, the waveform amplitude peaks near the “whirl” phase of the orbit and decreases during the “zoom” phase. During the early stage, when the particle is far from the black hole in the weak-field regime, the waveform exhibits a smooth, low-amplitude profile. As the particle approaches the strong-field region near the horizon, its trajectory becomes increasingly distorted, leading to sharp increases in both frequency and amplitude. The waveform reaches its maximum near the horizon, reflecting the intense curvature effects. As demonstrated in Fig.~\ref{120GWdifferenceE}, the parameter $\epsilon_1$ primarily influences the phase evolution of the waveform, while its impact on amplitude remains subdominant. For space-based detectors such as LISA, this phase shift is particularly important: EMRI signals remain coherent over $\sim 10^5$ orbital cycles in the years of observations \cite{Barack:2003fp}, allowing small phase deviations to accumulate into a measurable secular dephasing during the prolonged inspiral. Such phase evolution is the most sensitive observable in matched-filtering analyses of GWs, enabling EFTGR and Schwarzschild waveforms to be possibly distinguished in the observations of future space-based detectors.

In summary, our analysis reveals that EFTGR corrections, governed by the coupling parameter $\epsilon_1$, can significantly alter the dynamics of periodic orbits and the resulting GW signals in spherically symmetric static spacetimes. These results underscore the potential of GW astronomy to probe strong-field deviations from GR within a theoretically consistent and observationally accessible framework.

\section*{Acknowledgements}

This work is supported by the National Natural Science Foundation of China under Grants No. 12275238, No. 11675143, No. 12475056, No. 12247101, No. 12533001, No. 12542053 and No. 12473001, the National Key Research and Development Program under Grant No. 2020YFC2201503, and the Zhejiang Provincial Natural Science Foundation of China under Grants No. LR21A050001 and No. LY20A050002, and the Fundamental Research Funds for the Provincial Universities of Zhejiang in China under Grant No. RF-A2019015.

\end{document}